\documentclass[12pt,preprint]{aastex}
\usepackage{amsmath}

\begin{document}

\title{The Secular Evolution of the Primordial Kuiper
Belt\vspace*{0.8in}}

\author{Joseph M. Hahn}
\affil{Lunar and Planetary Institute, 3600 Bay Area Boulevard,
Houston, TX 77058\\
email: hahn@lpi.usra.edu\\
phone: 281--486--2113\\
fax: 281--486--2162\vspace{0.5in}\\
Submitted to {\it The Astrophysical Journal} April 5, 2003\\
Revised May 26, 2003\\
Accepted May 29, 2003\vspace{0.5in}
}

\begin{abstract}

A model that rapidly computes the secular 
evolution of a gravitating disk--planet system is developed. The
disk is treated as a nested set of gravitating
rings, with the rings'/planets'
time--evolution being governed by the classical
Laplace--Lagrange solution for secular evolution but
modified to account for the disk's finite
thickness $h$. The Lagrange planetary equations for this
system yield a particular class of spiral wave solutions,
usually denoted as apsidal density waves and nodal bending
waves. There are two varieties of apsidal waves---long waves and
short waves. Planets typically launch
long density waves at the disk's nearer edge or else
at a secular resonance in the disk, and these waves ultimately
reflect downstream at a more distant disk edge
or else at a $Q$--barrier in the disk, whereupon they
return as short density waves. Planets also launch nodal bending
waves, and these have the interesting property that they can
stall in the disk, that is, their group velocity plummets to
zero upon approaching a region in the disk that is too thick to
support the continued propagation of bending waves.

The rings model is used to compute the secular evolution of a
Kuiper Belt having a variety of masses, and it is
shown that the early massive Belt was very susceptible to
the propagation of low--amplitude apsidal and nodal waves launched
by the giant planets.
For instance, these waves typically excited orbits to
$e\sim\sin i\sim0.01$ in a primordial Kuiper Belt of mass
$M_{KB}\sim30$ Earth--masses. Although these orbital disturbances
are quite small, the resulting fractional variations
in the disk's surface density  due to the short density waves
is usually large, typically of order unity. This epoch of apsidal
and nodal wave propagation probably lasted throughout the
Belt's first $\sim10^7$ to $\sim5\times10^8$ years,
with the waves being shut off between the time when
the large $R\gtrsim100$ km KBOs first formed and when the
Belt was subsequently eroded and stirred up to its present
configuration.
\end{abstract}

\section{Introduction}
\label{introduction}

The Kuiper Belt is a vast swarm of comets orbiting at the
Solar System's outer edge. This Belt is comprised of debris
that was left over from the epoch of planet formation, and this
swarm's distribution of orbit elements preserves a record of
events that had occurred when the Solar System was still quite
young. A common goal of nearly all dynamical studies of the
Kuiper Belt is to decipher this Kuiper Belt record. However
this record is still open to some interpretation.
 
The dots in Fig.\ \ref{compare_ei}
give the Kuiper Belt Objects
(KBOs) eccentricities $e$ and inclinations $i$ versus
their semimajor axes $a$. This Figure reveals the KBOs' three
major dynamical classes: the Plutinos which inhabit Neptune's 3:2
resonance at $a=39.5$ AU, the Main Belt KBOs which are the
non--resonant KBOs orbiting between $40\lesssim a\lesssim48$
AU, and the more distant Scattered KBOs that live in eccentric,
nearly Neptune--crossing orbits. The Figure also shows that the
Plutinos and the Scattered KBOs have inclinations that span
$0\lesssim i\lesssim30^\circ$, while the Main Belt KBOs appear
to have a bimodal distribution of inclinations centered on
$i\simeq2^\circ$ and $i\simeq17^\circ$ \citep{B01}.
Note that accretion models show that these large $\sim100+$ km
KBOs must have first formed 
from much smaller planetesimal seeds that
were initially in nearly circular
and coplanar orbits having $e$ and $\sin i\lesssim0.001$
\citep{KL99}. However gravitational self--stirring cannot
account for the Kuiper Belt's current excited state, so one or
more mechanisms must also have stirred up the Kuiper Belt since
the time of formation.

The orbits of the Scattered KBOs are
perhaps the most easily understood. These objects likely have had
one or more  close encounters with Neptune which lofted these
bodies into eccentric, inclined orbits  \citep{DL97}. Repeated
encounters with Neptune cause these objects' semimajor axes
and eccentricities to evolve
stochastically along the Neptune--crossing curve shown
in Fig.\ \ref{compare_ei},
and most of these bodies are ultimately
ejected or accreted by the giant planets. However Neptune has
numerous weak, high--order mean motion resonances that thread
the Kuiper Belt, and these resonances permit some of these
Scattered objects to diffuse to lower eccentricities.
This allows a small percentage of the Scattered
objects to persist over the age of the Solar System at
eccentricities just below the Neptune--crossing curve seen
Fig.\ \ref{compare_ei} \citep{DL97}.
This diffusion to lower eccentricities
may also have been more pronounced had Neptune's
orbit also migrated outwards. In particular, \cite{G03} shows
that during the epoch of planet migration,
mean motion and secular resonances act as pathways that
allow some scattered KBOs to descend {\sl irreversibly} into
lower--eccentricity orbits that are far from the
Neptune--crossing curve and hence stable. Since 
Scattered KBOs have large inclinations of $i\gtrsim10^\circ$,
this process might also account for the Main Belt's bimodal
inclination distribution, with the $i\sim2^\circ$ component
representing the Main Belt's native population of
low--inclination KBOs, and the $i\sim17^\circ$ being due
to scattered invaders who were deposited in the Main Belt by
Neptune. However this process is quite inefficient since only
$\varepsilon\sim0.1\%$ of these Scattered KBOs manage to find
orbits that are stable over a solar age \citep{G03}.

The possibility that Neptune's orbit had expanded outwards is
also supported by the cluster of KBOs that inhabit Neptune's 3:2
resonance (see Figure \ref{compare_ei}).
This may have occurred when Neptune had first formed and
began to vigorously scatter the local planetesimal debris.
This process can drive an exchange of angular momentum between
the planets and the planetesimal disk \citep{FI84}, so this
episode of disk--clearing can result in a rearrangement of all
of the giant planets' orbits over a timescale of $\sim10^7$ years
\citep{HM99}. Outward planet migration will also cause Neptune's
mean--motion resonances to sweep out across the primordial Kuiper
Belt, and these migrating resonances are quite
effective at capturing KBOs and pumping up their eccentricities
\citep{M95}. Models of this process show that if Neptune's orbit
had in fact smoothly expanded some $\Delta a\sim7$ AU over a
timescale longer than a few million years, then resonance
capture would have deposited numerous KBOs in the 3:2 and the
2:1 resonances with eccentricities
distributed over $0\lesssim e\lesssim0.3$
\citep{M95, CJ02}. Although numerous KBOs do indeed inhabit
Neptune's 3:2 resonance, only a handful are known to live near
the 2:1 resonance at $a=47.8$ AU, and many of these bodies have
eccentricities of $e\sim0.3$ which puts them quite near the
Neptune--crossing curve (see Fig.\ \ref{compare_ei}). Thus it is
possible that some or all of the bodies orbiting near $a=47.8$ AU
might instead be members of the Scattered Belt. Thus if planet
migration did indeed occur, then the apparent low abundance of
KBOs at the 2:1 resonance is a mystery that can only be partly
due to the observational bias that selects against the
discovery of lower eccentricity objects at the 2:1 \citep{JLT98}.

But of particular interest here is the Main Belt which
preserves additional evidence for another mechanism having
stirred up the Kuiper Belt. Again, if Neptune
did indeed migrate outwards $\Delta a\sim7$ AU, then the entire
Main Belt was swept by the advancing 2:1 resonance.
Models of planet migration show that the efficiency of resonance
capture is no more than $\sim50\%$ \citep{CJ02}, so the current
members of the Main Belt evidently avoided permanent capture by
slipping through the advancing 2:1 resonance.
However this planet migration
scenario is utterly unable to account for the high
inclinations of $0\lesssim i\lesssim30^\circ$ observed in the
Main Belt (see Fig.\ \ref{compare_ei}), since the N--body
simulations show that the advancing 2:1 resonance typically
excites Main Belt
inclinations by only a few degrees \citep{M95, CJ02}. Evidently,
an additional mechanism is also responsible for exciting the
Main KBO Belt. It has been shown that sizable KBO excitation can
occur if a recently--formed Neptune had been scattered outwards
by  Jupiter and/or Saturn into a Belt--crossing orbit
\citep{TDL02}.  Another possible source of KBO excitation is the
invasion of the Main Belt by the Scattered Belt \citep{G03}.
However this latter process is a very inefficient mechanism
having  $\varepsilon\sim0.001$; an alternate mechanism that is
possibly more efficient will be explored below.

It has also been suggested that secular resonance
sweeping may have been responsible for exciting the Kuiper Belt
\citep{NI00}. Note that the locations of the secular resonances
are very sensitive to the Solar System's mass distribution.
Consequently, the depletion of the solar nebula (which includes
perhaps $\sim99\%$ of the Solar System's initial mass content)
could have driven these secular resonances across vast tracts of
the Solar System. Indeed, the model by \cite{NI00} suggests that
if the nebula was depleted on a timescale of $\tau\sim10^7$
years, Main Belt inclinations of $i\sim20^\circ$ can be excited
due to the passage of the $\nu_{15}$ secular resonance that
sweeps outwards to infinity as the nebula is depleted\footnote{It
should be noted that the resulting inclination excitation is
also sensitive to the tilt between the nebula midplane and the
invariable plane. For instance \cite{NI00} place the nebula
midplane in the ecliptic, and this
results in substantial excitation.
But if the nebula midplane is instead placed in the invariable
plane, which is tilted $1.6^\circ$ from the ecliptic,
then almost no excess excitation results \citep{HW02}. We also
note that the nebula models of \cite{NI00} as well as \cite{HW02}
both treat the gas disk as a rigid slab of gas. However a more
realistic treatment would allow the nebula disk to flex and warp
in response to the planets' secular perturbations.
It is suspected that this additional degree of freedom will
substantially alter the secular resonance sweeping; indeed, it
can be argued that the $\nu_{15}$ never did sweep across the
Kuiper Belt on account of this flexure (E.\ Chiang and W.\ Ward,
2002, private communication), so perhaps secular resonance
sweeping of the Kuiper Belt
is actually a moot issue.}. However this
model is rather idealized in that it treats the Kuiper Belt as
massless, which is a concern since a primordial
Kuiper Belt having some mass is also susceptible to the
propagation of very long--wavelength spiral waves that
can be launched at a secular resonance in the disk
\citep{WH98aj, WH03}. This issue is worth further examination
since wave--action can alter the magnitude of resonant excitation
considerably \citep{HWR95, WH98aj}. But even if
there is no resonance in the disk, planets orbiting interior to
a particle disk can still launch these spiral waves at the
disk's inner edge \citep{WH98sci}. Preliminary results from a
model of secular resonance sweeping also reveals that a Kuiper
Belt having only a modest amount of mass is utterly awash in
these waves once the nebula is depleted \citep{HW02}. However
the purpose of the present study is to
first characterize the properties of these waves in the simpler,
post--nebula environment, and to explore their cosmogonic
implications. Thus the following will consider
a suite of models of the secular evolution of the outer Solar
System for primordial Kuiper Belts having a variety of masses.

Accretion models tell us that the primordial Kuiper Belt must
have had a mass of
$M_{\mbox{\scriptsize KB}}\sim30$ M$_\oplus$ in the $30<a<50$ AU
interval in order for Pluto and its cohort of KBOs to have formed
and survive over the age of the Solar System \citep{KL99}. A
similar Kuiper Belt mass is also needed to drive Neptune's
orbital migration of $\Delta a\sim7$ AU \citep{HM99}. However
the current mass is
$M_{\mbox{\scriptsize KB}}\sim0.2$ M$_\oplus$ \citep{JLT98}, so
the Kuiper Belt appears to have been eroded by a factor of
$\sim150$. This may be due to a dynamical erosion of the Belt by
Neptune or possibly by other perturbers that may once have been
roaming about the outer Solar System, as well as due to the
collisional erosion that has since ground all of the smaller KBOs
down to dust grains that are then removed from the Solar System
by radiation forces \citep{KL99, KB01}.
The following will give results
obtained from models of the secular evolution of the outer Solar
System for Kuiper Belts having masses in the interval
$0\le M_{\mbox{\scriptsize KB}}\le30$ M$_\oplus$.

Section \ref{model} derives the so--called rings model that
will be used to study the secular evolution of disk--planet
systems; the reader uninterested in these details might skip
ahead to Section \ref{waves} or \ref{results}.
Since spiral density and bending waves appear prominently in the
model results, their properties are examined in
Section \ref{waves}. Section \ref{results} will describe
the model's application to the primordial Kuiper Belt, and a
summary of results is then given in Section \ref{summary}.

\section{The Secular Evolution of Disk--Planet Systems}
\label{model}

The following shall treat the disk as a collection of nested
gravitating rings in orbit about the Sun. Their mutual
perturbations will cause these rings' to slowly flex and tilt
over time, and this evolution is governed by the Lagrange
planetary equations. 

\subsection{the rings model}
\label{rings model}

Begin with the gravitational potential that a single perturbing
ring of mass $m'$ exerts at the point $\bf{r}$ on another ring
of mass $m$:
\begin{equation}
\label{Phi'1}
\Phi'(\mbox{\bf{r}})=-\int\frac{G\rho'dV'}{\Delta}
\end{equation}
where $G$ is the gravitation constant, $\rho'$ is the mass
density of the differential volume element $dV'$,
$\Delta$ is the separation between the
perturbing mass element $\rho'dV'$ at $\bf{r'}$ and the field
point $\bf{r}$, and the integration proceeds over the
three dimensional extent of ring $m'$.
In the following, primed quantities will refer to the perturbing
ring $m'$ and unprimed quantities will refer
to the perturbed ring $m$. 
Each ring can be thought of as a swarm of
numerous particles all having a common semimajor axis $a$
and an identical mean orbital eccentricity $e$, inclination
$i$, longitude of periapse $\tilde{\omega}$, and longitude of
ascending node $\Omega$. It is also assumed that these
particles have an isotropic dispersion velocity $c$ that 
gives rise to the ring's finite radial as well as
vertical half--thickness
$h\simeq c/n$, where $n$ is the ring's mean motion. It shall also
be assumed that the density $\rho'$ varies only in the
azimuthal direction due to the keplerian motion of
the ring's particles; in this case the density is
$\rho'=\lambda'/4{h'}^2$
where
\begin{equation}
\lambda'=\frac{m'r'}{2\pi {a'}^2\sqrt{1-{e'}^2}}
\end{equation}
is the ring's azimuthal mass per unit length \citep{MD99}. 
In cylindrical coordinates $\bf{r'}$$=(\ell', \phi', z')$ 
and $dV'=\ell'd\ell'd\phi'dz'$. If the slight radial variations
in equation (\ref{Phi'1})'s integrand are ignored
({\it i.e.}, $\int\ell'd\ell'\simeq2r'h'$),
the potential becomes
\begin{equation}
\Phi'(\mbox{\bf{r}})\simeq-\int_{-\pi}^{\pi}d\phi'
  \int_{z_o'-h'}^{z_o'+h'}dz'\frac{G\lambda'r'}{2h'\Delta}.
\end{equation}
where $z_o'(\phi')$ is the longitude--dependent height of the
perturbing ring's midplane from the $z=0$ plane. Of course the
perturbed ring $m$ also has a radial and vertical
half--thickness $h$, and it is useful to form an effective
potential by averaging $\Phi'(\mbox{\bf{r}})$ over the radial
and vertical extent of ring $m$:
\begin{equation}
\label{Phi'2}
<\Phi'(\mbox{\bf{r}})>=\int_{-h}^{h}\frac{d\ell}{2h}
\int_{z_o-h}^{z_o+h}\frac{dz}{2h}\Phi'(\mbox{\bf{r}})
\equiv-\int_{-\pi}^{\pi}d\phi'\frac{G\lambda'r'}{r}Q
\end{equation}
where
\begin{equation}
Q\simeq\int_{z_o-h}^{z_o+h}\frac{dz}{2h}
  \int_{z_o'-h'}^{z_o'+h'}\frac{dz'}{2h'}\frac{r}{\Delta}
\end{equation}
where again the slight variations in the integrand with
$\ell$ are ignored, and that only $\Delta$ is assumed to be
sensitive to the variations in $z$ and $z'$. Note that this
averaging of $\Phi'$ is essential in order for the algorithm
developed below to conserve angular momentum.

The next task is to evaluate the double integrals in $Q$.
First note that the separation $\Delta$
between the perturbing mass element
$\rho'dV'$ at $\bf{r'}$ and the field point $\bf{r}$ obeys
$\Delta^2=r^2+{r'}^2-2\sqrt{(r^2-z^2)({r'}^2-{z'}^2)}
\cos(\phi'-\phi)
-2zz'$ where $r^2=\ell^2+z^2$. Setting $\alpha\equiv r'/r$,
$\beta\equiv z/r$, and $\beta'\equiv z'/r'$, then 
\begin{equation}
\left(\frac{\Delta}{r}\right)^2\simeq1+\alpha^2
  -2\alpha\cos(\phi'-\phi)
  +(\beta^2+{\beta'}^2)\alpha\cos(\phi'-\phi)-2\alpha\beta\beta'
\end{equation}
to second order in the $\beta$'s, which are of order the rings'
inclinations and are assumed small. Inserting this into the
expression for $Q$ yields
\begin{multline}
\label{Q1}
Q=\frac{1}{4\mathfrak{h}\mathfrak{h}'}
\int_{\beta_o-\mathfrak{h}}^{\beta_o+\mathfrak{h}}d\beta
\int_{\beta_o'-\mathfrak{h'}}^{\beta_o'+\mathfrak{h'}}
  d\beta'\left[1+\alpha^2-2\alpha\cos\Delta\phi
    +(\beta^2+{\beta'}^2)\alpha\cos\Delta\phi
    -2\alpha\beta\beta'\right]^{-1/2}\\
=\int_{\beta_o'-\mathfrak{h'}}^{\beta_o'+\mathfrak{h'}}
  \frac{\ln(\Upsilon)d\beta'}
  {4\mathfrak{h}\mathfrak{h}'\sqrt{\alpha\cos\Delta\phi}}
\end{multline}
where $\mathfrak{h}\equiv h/r\simeq h/a$ and
$\mathfrak{h'}\equiv h'/r'\simeq h'/a'$ are the fractional
half--thicknesses of rings $m$ and $m'$,
$\beta_o\equiv z_o/r$ and $\beta_o'\equiv z_o'/r'$
are the rings' midplane latitudes, $\Delta\phi\equiv\phi'-\phi$,
and the right--hand side of equation (\ref{Q1}) is the result of
doing the integration in $\beta$, where 
\begin{equation}
\Upsilon=\frac{\beta'\alpha-\beta_o\alpha\cos\Delta\phi-
  \mathfrak{h}\alpha\cos\Delta\phi-
  \sqrt{\alpha\cos\Delta\phi(\Gamma+\varepsilon)}}
{\beta'\alpha-\beta_o\alpha\cos\Delta\phi+
  \mathfrak{h}\alpha\cos\Delta\phi-
  \sqrt{\alpha\cos\Delta\phi(\Gamma-\varepsilon)}}
\end{equation}
with $\Gamma=1+\alpha^2-2\alpha\cos\Delta\phi+
(\beta_o^2+\mathfrak{h}^2+{\beta'}^2)\alpha\cos\Delta\phi
-2\alpha\beta'\beta_o$ and
$\varepsilon=2\mathfrak{h}(\beta_o\cos\Delta\phi-\beta')\alpha$.
The $\beta$'s and the $\mathfrak{h}$'s are assumed small, so
$\varepsilon$ is second order in the small quantities and is
negligible when compared to other terms. Thus
\begin{equation}
\Upsilon\simeq\frac{1-
  [\beta'\alpha-\beta_o\alpha\cos\Delta\phi-
    \mathfrak{h}\alpha\cos\Delta\phi]/
    \sqrt{\alpha\cos\Delta\phi\Gamma}}
{1-[\beta'\alpha-\beta_o\alpha\cos\Delta\phi+
    \mathfrak{h}\alpha\cos\Delta\phi]/
    \sqrt{\alpha\cos\Delta\phi\Gamma}}
\simeq1+2\mathfrak{h}\sqrt{\frac{\alpha\cos\Delta\phi}{\Gamma}}
\end{equation}
so $\ln\Upsilon\simeq2\mathfrak{h}
\sqrt{\alpha\cos\Delta\phi/\Gamma}$. This is inserted back into
equation (\ref{Q1}) and the remaining integral over
$\beta'$ is evaluated similarly, yielding
$Q\simeq\ln(\Lambda)/2\mathfrak{h}'\sqrt{\alpha\cos\Delta\phi}$
where
\begin{equation}
\Lambda=\frac{1-[\beta_o\alpha-\beta_o'\alpha\cos\Delta\phi-
  \mathfrak{h'}\alpha\cos\Delta\phi]/
    \sqrt{\alpha\cos\Delta\phi(\Psi+{\cal Z}+\xi)}}
{1-[\beta_o\alpha-\beta_o'\alpha\cos\Delta\phi+
  \mathfrak{h'}\alpha\cos\Delta\phi]/
    \sqrt{\alpha\cos\Delta\phi(\Psi+{\cal Z}-\xi)}}
\simeq1+2\mathfrak{h'}\sqrt{\frac{\alpha\cos\Delta\phi}
  {\Psi+{\cal Z}}}
\end{equation}
with $\Psi\equiv1+\alpha^2-2\alpha\cos\Delta\phi(1-H^2)\simeq
  (1+\alpha^2)(1+H^2)-2\alpha\cos\Delta\phi$,
$H^2\equiv(\mathfrak{h}^2+\mathfrak{h'}^2)/2$, 
${\cal Z}=(\beta_o^2+{\beta_o'}^2)\alpha\cos\Delta\phi
  -2\alpha\beta_o\beta_o'$ and 
$\xi\equiv2\mathfrak{h'}(\beta_o'\cos\Delta\phi-\beta_o)\alpha$
is another negligible term. Inserting
$\ln\Lambda\simeq2\mathfrak{h}'\sqrt{\alpha\cos\Delta\phi/
(\Psi+{\cal Z})}$ back into $Q$ yields
\begin{equation}
Q\simeq\frac{1}{\sqrt{\Psi+{\cal Z}}}
\simeq\Psi^{-1/2}-\frac{1}{2}{\cal Z}\Psi^{-3/2}.
\end{equation}

A Fourier expansion of $\Psi^{-s}$ will be useful, {\it i.e.},
$\Psi^{-s}=\onehalf\sum_{m=-\infty}^{\infty}
\tilde{b}^{(m)}_s\cos m(\phi'-\phi)$ where
\begin{equation}
\label{b_soft_int}
\tilde{b}^{(m)}_{s}(\alpha,\mathfrak{h},\mathfrak{h'})=
  \frac{2}{\pi}\int_0^\pi
\frac{\cos(m\phi)d\phi}{\{(1+\alpha^2)
[1+\onehalf(\mathfrak{h}^2+\mathfrak{h'}^2)-2\alpha\cos\phi\}^s}
\end{equation}
is the softened Laplace coefficient. The usual unsoftened form
is $b^{(m)}_{s}(\alpha)=\tilde{b}^{(m)}_{s}(\alpha,0,0)$. These
two coefficients are nearly equal when $\alpha$ is far from
unity, but the softened form is finite at $\alpha=1$
whereas the unsoftened form diverges. Writing $Q$ in terms of
softened Laplace coefficients thus gives
\begin{equation}
\label{Q2}
Q\simeq\frac{1}{2}\sum_{m=-\infty}^\infty\cos m(\phi'-\phi)
\left\{\tilde{b}^{(m)}_{1/2}-
\left[\frac{1}{2}(\beta_o^2+{\beta_o'}^2)\alpha\cos\Delta\phi
  -\beta_o\beta_o'\alpha\right]\tilde{b}^{(m)}_{3/2}\right\},
\end{equation}
which is then inserted in equation (\ref{Phi'2}) to get
the perturbing ring's potential
\begin{equation}
\label{Phi'3}
<\Phi'>=-\int_{-\pi}^\pi d\phi'\frac{G\lambda'r'}{2r}
\sum_{m=-\infty}^{\infty}\cos m(\phi'-\phi)
\left\{\tilde{b}^{(m)}_{1/2}-
\left[\frac{1}{2}(\beta_o^2+{\beta_o'}^2)\alpha\cos(\phi'-\phi)
-\alpha\beta_o\beta_o'\right]
\tilde{b}^{(m)}_{3/2}\right\}.
\end{equation}

The final task of this Section is to write the
rings' coordinates in terms of its orbit elements
\begin{subequations}
\label{orbits}
\begin{align}
r\simeq&a(1-e\cos\nu-\frac{1}{2}e^2+\frac{1}{2}e^2\cos2\nu)\\
\phi=&\tilde{\omega}+\nu\simeq\tilde{\omega}+M+2e\sin M\\
\beta_o=&\frac{z_o}{r}\simeq\sin i\sin(\phi-\Omega)
\end{align}
\end{subequations}
where $\nu$ is the true anomaly of a ring--element at $\bf{r}$
and $M=nt$ is the corresponding mean anomaly. Inserting
equations (\ref{orbits}) into the potential $<\Phi'>$,
expanding to second order in the $e$'s and $i$'s, doing the
$\phi'$ integration in equation (\ref{Phi'3}), and
time--averaging the resulting expression over the orbital period
of ring $m$ then yields the time--averaged
potential $<\bar{\Phi}'>$ experienced by ring $m$ due to ring
$m'$ assuming small eccentricities and inclinations:
\begin{equation}
\label{Phi'4}
\begin{split}
<\bar{\Phi}'>=-\frac{Gm'}{a}\left[
\frac{1}{2}\tilde{b}^{(0)}_{1/2}+
\frac{1}{8}(e^2+{e'}^2)f
+\frac{1}{4}ee'\cos(\tilde{\omega}'-\tilde{\omega})g
\right. \\
\left.-\frac{1}{8}(i^2+{i'}^2)\alpha\tilde{b}^{(1)}_{3/2}
+\frac{1}{4}ii'\cos(\Omega'-\Omega)
\alpha\tilde{b}^{(1)}_{3/2}\right]
\end{split}
\end{equation}
where the $f$ and $g$ functions are
\begin{subequations}
\label{fg}
\begin{align}
\label{f}
f(\alpha,\mathfrak{h},\mathfrak{h'})&=
\left(2\alpha\frac{\partial}{\partial\alpha}
+\alpha^2\frac{\partial^2}{\partial\alpha^2}
\right)b^{(0)}_{1/2}
=\alpha\tilde{b}^{(1)}_{3/2}-
  3\alpha^2H^2(2+H^2)\tilde{b}^{(0)}_{5/2}\\
\label{g}
g(\alpha,\mathfrak{h},\mathfrak{h'})&=
\left(2-2\alpha\frac{\partial}{\partial\alpha}
-\alpha^2\frac{\partial^2}{\partial\alpha^2}
\right)b^{(1)}_{1/2}
=-\alpha\tilde{b}^{(2)}_{3/2}+
  3\alpha^2H^2(2+H^2)\tilde{b}^{(1)}_{5/2}
\end{align}
\end{subequations}
where $H^2=\onehalf(\mathfrak{h}^2+\mathfrak{h'}^2)$
and $\alpha$ has been redefined as $\alpha=a'/a$.
The right--hand side of equations (\ref{fg}) are derived in
Appendix \ref{appendixB}, and it is shown in
Appendix \ref{appendixC}
that the softened Laplace coefficients can be
written in terms of complete elliptic integrals. Consequently,
functions $f$, $g$, and $\tilde{b}^{(m)}_{s}$ all can be 
rapidly evaluated without relying upon a numerical integration
of equation (\ref{b_soft_int}).
It is also noted that the $\tilde{b}^{(m)}_{s}$, $f$, and $g$
functions obey the following reciprocal relations
\begin{subequations}
\label{reciprocal}
\begin{align}
\label{reciprocal b}
\tilde{b}^{(m)}_{s}(\alpha^{-1},\mathfrak{h'},\mathfrak{h})=&
  \alpha^{2s}
  \tilde{b}^{(m)}_{s}(\alpha,\mathfrak{h},\mathfrak{h'})\\
  f(\alpha^{-1},\mathfrak{h'},\mathfrak{h})=&
    \alpha f(\alpha,\mathfrak{h},\mathfrak{h'})\\
\label{g_reciprocal}
  g(\alpha^{-1},\mathfrak{h'},\mathfrak{h})=&
    \alpha g(\alpha,\mathfrak{h},\mathfrak{h'}).
\end{align}
\end{subequations}
These relations will be used in Section \ref{Lconservation} to
show that the equations of motion developed below
conserve angular momentum.

The laborious procedure of expanding, integrating, and
then time--averaging $<\Phi'>$ is not included here
since a similar analysis can be found in
\cite{MD99}\footnote{Actually \cite{MD99} derive the
time--averaged acceleration (rather than a potential) that ring
$m'$ exerts on $m$, which they insert into the Gauss equations
to obtain a set of dynamical equations equivalent to that
obtained here when $\mathfrak{h}=0$.}. Since the terms
proportional to ${e'}^2$ and ${i'}^2$ as well as the first
term in equation (\ref{Phi'4}) do not contribute to the resulting
dynamical equations, they may be neglected. The 
disturbing function $R$ for ring $m$ due to another
ring $m'$ is $-1\times$ the surviving terms in
$<\bar{\Phi}'>$, {\it i.e.},
\begin{equation}
\label{R_soft}
R=\frac{Gm'}{a}\left[\frac{1}{8}fe^2
+\frac{1}{4}gee'\cos(\tilde{\omega}'-\tilde{\omega})
-\frac{1}{8}\alpha\tilde{b}^{(1)}_{3/2}i^2
+\frac{1}{4}\alpha\tilde{b}^{(1)}_{3/2}
ii'\cos(\Omega'-\Omega)\right].
\end{equation}
Note that when $\mathfrak{h}=\mathfrak{h'}=0$
the disturbing function for a point mass
$m$ perturbed by point mass $m'$ is recovered
\citep{BC61, MD99}, which is to be expected since both a point
mass as well as a thin ring have the same disturbing function
to this degree of accuracy \citep{MD99}.

It should also be noted that many celestial mechanics texts
develop two distinct expressions for the disturbing
function $R$, one due to a perturber in an interior orbit
having $a'<a$, and another expression due to an exterior
perturber having $a'>a$ [c.f., \cite{BC61, MD99}]. However this
pairwise development is unnecessary in this application since
equation (\ref{R_soft}) is valid for
$\alpha=a'/a<1$ as well as for $\alpha>1$. Indeed, it is
straightforward to show that these pairs of disturbing functions,
such as equations (7.6) and (7.7) in \cite{MD99}, are in fact
equivalent to equation (\ref{R_soft}) with
$\mathfrak{h}=\mathfrak{h'}=0$; they only appear
distinct since one is a function of $\alpha$ and the other
is actually a function of $\alpha^{-1}$.

In terms of the variables
\begin{subequations}
\begin{align}
h=&e\sin\tilde{\omega}& p=&i\sin\Omega \\
k=&e\cos\tilde{\omega}& q=&i\cos\Omega,
\end{align}
\end{subequations}
the disturbing function can then be written
\begin{equation}
\label{R_hk}
\begin{split}
R=n^2a^2\left(\frac{m'}{M_\odot+m}\right)
\left[\frac{1}{8}f(h^2+k^2)
+\frac{1}{4}g(hh'+kk')
-\frac{1}{8}\alpha
\tilde{b}^{(1)}_{3/2}(p^2+q^2)\right.+\\
\left.\frac{1}{4}\alpha
\tilde{b}^{(1)}_{3/2}(pp'+qq')\right]
\end{split}
\end{equation}
where the mean motion $n=\sqrt{G(M_\odot+m)/a^3}$ with $M_\odot$
being the solar mass.

\subsubsection{the N ring problem}
\label{N rings}

For the more general problem of $N$ perturbing rings, replace the
perturbing ring mass $m'$ with $m_k$ and give all the other
primed quantities the subscript $k$. The disturbing
function $R\rightarrow R_j$ for the perturbed ring having a mass
$m\rightarrow m_j$ is sum of equation (\ref{R_hk}) over all
the other rings $k\ne j$. Setting $\alpha_{jk}\equiv a_k/a_j$,
$n_j=\sqrt{G(M_\odot+m_j)/a_j^3}$, and
\begin{subequations}
\begin{align}
A_{jj}&=\frac{n_j}{4}\sum_{k\ne j}
  \left(\frac{m_k}{M_\odot+m_j}\right)
  f(\alpha_{jk},\mathfrak{h}_j,\mathfrak{h}_k)\\
A_{jk}&=\frac{n_j}{4}\left(\frac{m_k}{M_\odot+m_j}\right)
  g(\alpha_{jk},\mathfrak{h}_j,\mathfrak{h}_k)
  \quad\mbox{\hspace{10ex}$j\ne k$}\quad\\
B_{jj}&=-\frac{n_j}{4}\sum_{k\ne j}
  \left(\frac{m_k}{M_\odot+m_j}\right)
  \alpha_{jk}\tilde{b}^{(1)}_{3/2}
  (\alpha_{jk},\mathfrak{h}_j,\mathfrak{h}_k)\\
B_{jk}&=\frac{n_j}{4}\left(\frac{m_k}{M_\odot+m_j}\right)
  \alpha_{jk}\tilde{b}^{(1)}_{3/2}
  (\alpha_{jk},\mathfrak{h}_j,\mathfrak{h}_k)
  \quad\mbox{\hspace{5ex}$j\ne k$,}\quad
\end{align}
\end{subequations}
that sum becomes
\begin{equation}
\label{R_j}
R_j=n_ja_j^2\left[\frac{1}{2}A_{jj}(h_j^2+k_j^2)
+\sum_{k\ne j}A_{jk}(h_jh_k+k_jk_k)
+\frac{1}{2}B_{jj}(p_j^2+q_j^2)
+\sum_{k\ne j}B_{jk}(p_jp_k+q_jq_k)\right]
\end{equation}
in the notation of \cite{MD99}. The $A_{jk}$ and the
$B_{jk}$ can be regarded as two $N\times N$ matrices $\bf{A}$
and $\bf{B}$ whose entries describe the magnitude of the
mutual gravitational interactions that are exerted among the $N$
rings. In the following, quantities having a $j$ subscript will
always refer to the perturbed ring in question while the $k$
subscript will always refer to another perturbing ring.

The time--variation of the rings' orbit elements is given by
the Lagrange planetary equations; to lowest order in the $e$'s
and $i$'s these are \citep{BC61, MD99}
\begin{subequations}
\label{hkpq_dot}
\begin{align}
\frac{dh_j}{dt}\simeq&\frac{\partial R_j/\partial k_j}{n_ja_j^2}
  =\sum_{k=1}^NA_{jk}k_k\\
\frac{dk_j}{dt}\simeq&-\frac{\partial R_j/\partial h_j}{n_ja_j^2}
  =-\sum_{k=1}^NA_{jk}h_k\\
\frac{dp_j}{dt}\simeq&\frac{\partial R_j/\partial q_j}{n_ja_j^2}
  =\sum_{k=1}^NB_{jk}q_k\\
\frac{dq_j}{dt}\simeq&-\frac{\partial R_j/\partial p_j}{n_ja_j^2}
  =-\sum_{k=1}^NB_{jk}p_k,
\end{align}
\end{subequations}
and their well--known Laplace--Lagrange solution is
\begin{subequations}
\label{solution}
\begin{align}
h_j(t)=&\sum_{i=1}^NE_{ji}\sin(g_it+\beta_i)\\
k_j(t)=&\sum_{i=1}^NE_{ji}\cos(g_it+\beta_i)\\
p_j(t)=&\sum_{i=1}^NI_{ji}\sin(f_it+\gamma_i)\\
q_j(t)=&\sum_{i=1}^NI_{ji}\cos(f_it+\gamma_i)
\end{align}
\end{subequations}
where $g_i$ is the $i^{\mbox{\scriptsize th}}$ eigenvalue of the
$\bf{A}$ matrix, $f_i$ is the $i^{\mbox{\scriptsize th}}$
eigenvalue of $\bf{B}$, $E_{ji}$ is the $N\times N$
matrix formed from the $N$ eigenvectors to $\bf{A}$ and $I_{ji}$
is the matrix of eigenvectors to $\bf{B}$, and $\beta_i$ and
$\gamma_i$ are integration constants. 

To apply this rings model, first assign to the $N$ rings their
masses $m_j$, their semimajor axes $a_j$,
and their fractional half--thickness $\mathfrak{h}_j$.
Planets are represented as thin rings having $\mathfrak{h}_j=0$.
Then construct the system's
$\bf{A}$ and $\bf{B}$ matrices and compute
their eigenvalues $g_i$, $f_i$ and the eigenvector
arrays $E_{ji}$ and $I_{ji}$. The rings' initial
orbits $e_j(0), i_j(0), \tilde{\omega}_j(0),$ and $\Omega_j(0)$
are then used to determine the integration
constants $\beta_i$ and $\gamma_i$ as well as to rescale the
eigenvectors $E_{ji}$ and $I_{ji}$ such that equations
(\ref{solution}) agree with the initial conditions.
A handy recipe for this particular task is given in \cite{MD99}.
Equations (\ref{solution}) are then used to compute the system's
time--history, and orbit elements are recovered via
$e_j^2=h_j^2+k_j^2$, $i_j^2=p_j^2+q_j^2$,
$\tan\tilde{\omega}_j=h_j/k_j$, and $\tan\Omega_j=p_j/q_j$.

The rather laborious derivation given above thus confirms
the assertion by \cite{T01}
that one only needs to soften the Laplace coefficients in order
to use the Laplace--Lagrange solution for calculating the
secular evolution of a continuous disk. However Section
\ref{Lconservation} will show that this softening must be done
judiciously such that equation (\ref{reciprocal b}) is obeyed
in order for the solution to preserve the system's angular
momentum.

\subsection{tests of the rings model}
\label{tests}

Several tests have been devised in order to demonstrate that
the rings model described above behaves as expected.

\subsubsection{angular momentum conservation}
\label{Lconservation}

The equations developed above conserve the system's total
$z$--component of angular momentum
$L_z=\sum_jm_jn_ja_j^2\sqrt{1-e_j^2}\cos(i_j)$. To show this,
expand $L_z$ to second order in the $e's$ and
$i's$, which is the same degree of precision to which the
disk potential is developed above. This gives
$L_z\simeq L_0-L_e-L_i$ where
\begin{subequations}
\label{Lei}
\begin{align}
\label{Le}
L_e=&\frac{1}{2}\sum_jm_jn_ja_j^2e_j^2\\
L_i=&\frac{1}{2}\sum_jm_jn_ja_j^2i_j^2
\end{align}
\end{subequations}
and $L_0=\sum_jm_jn_ja_j^2$ is a constant. It is shown in
Appendix \ref{appendixD} that the dynamical equations
(\ref{hkpq_dot}) conserve angular momenta to second order in the
$e$'s  and $i$'s, that is, $dL_e/dt=0$ and $dL_i/dt=0$. When the
rings model is used to calculate the
secular evolution of a `sparse' system, such as the
Jupiter--Saturn system described below or one composed of all
four giant planets, the angular
momenta $L_e$ and $L_i$ are preserved to near machine limits
to a fractional precision of $\sim10^{-7}$ in this
single floating--point implementation. This is true regardless
of whether the planets are represented as thin rings having
$\mathfrak{h}_j=0$ or as thick rings with $\mathfrak{h}_j>0$.
However $L_z$
conservation is a little worse, $\sim10^{-6}$, in the `crowded'
systems described in Section \ref{results} that
consist of a few planets plus a disk comprised of many
closely--packed rings. However these errors are always smaller,
by a factor of $\sim10^4$ to $10^5$, than the angular momenta
associated with the disturbances and waves seen in these disks.
Thus the wave--like disk behavior reported below is
real and is not due to a diffusion of some numerical error.

\subsubsection{Jupiter and Saturn}
\label{JS}

\cite{MD99} use the Laplace--Lagrange planetary solution, 
equations (\ref{solution}),
to study a system composed of Jupiter and Saturn. The rings
model developed here reproduces that system's
$\bf{A}$ and $\bf{B}$ matrices, eigenvector
arrays $E_{ji}$ and $I_{ji}$, eigenvalues $g_i$ and $f_i$,
and integration constants $\beta_i$ and $\gamma_i$, to the
same precision quoted by \cite{MD99}, provided one sets
$n_j=\sqrt{GM_\odot/a_j^3}$ and $\mathfrak{h}_j=0$.
The rings model also reproduces the
figures in \cite{MD99} that show this system's orbital history,
as well as the figures showing the forced orbit elements of
numerous other massless `test--rings' orbiting throughout
this system. However is should be noted that rigorous angular
momentum conservation actually requires setting
$n_j=\sqrt{G(M_\odot+m_j)/a_j^3}$ instead.

\subsubsection{precession in an axisymmetric disk}
\label{axisymmetric disk}

\cite{H80} points out that a massless test particle orbiting in
a smooth, axisymmetric disk experiences a {\sl regression} of its
longitude of periapse,
{\it i.e.}, $\dot{\tilde{\omega}}<0$, which is
the opposite of the usual prograde apse precession that occurs
throughout the Solar System. As \cite{W81} shows, it is the
nearby disk parcels whose orbits actually cross the test
particle's orbit that drives periapse regression at a rate that
exceeds the prograde contribution from the more distant parts of
the disk.

The rings code also reproduces this phenomenon. An annulus having
a surface density $\sigma'$, mass
$\delta m'=2\pi\sigma'a'da'$, a semimajor axis $a'$, and a
fractional half--thickness $\mathfrak{h'}$ contributes
$\delta R=n^2a^2[f(\alpha,0,\mathfrak{h'})e^2-
  \alpha\tilde{b}^{(1)}_{3/2}(\alpha,0,\mathfrak{h'})i^2]
  \delta m'/8M_\odot$
to the particle's disturbing function [see equation
(\ref{R_soft}) with $e'=i'=0$]. Adopting a power--law in the disk
surface density, $\sigma'=\sigma(a)\alpha^{-r}$ where
$\alpha=a'/a$, the total disturbing function integrated across
a semi--infinite disk is
\begin{equation}
R=\int\delta R=\frac{1}{2}\mu_dn^2a^2
  \left[I_{\tilde{\omega}}e^2-I_{\Omega}i^2\right]
\end{equation}
where $\mu_d=\pi\sigma(a)a^2/M_\odot=\pi G\sigma/an^2$
is the so--called normalized disk mass and 
\begin{subequations}
\label{II}
\begin{align}
\label{Iomega}
I_{\tilde{\omega}}=&\frac{1}{2}\int_0^\infty\alpha^{1-r}
  f(\alpha,0,\mathfrak{h'})d\alpha\\
\label{IOmega}
I_{\Omega}=&\frac{1}{2}\int_0^\infty\alpha^{2-r}
  \tilde{b}^{(1)}_{3/2}(\alpha,0,\mathfrak{h'})d\alpha.
\end{align}
\end{subequations}
Note that $I_{\tilde{\omega}}$ and $I_{\Omega}$ are
double integrals according to the definitions of
$\tilde{b}^{(m)}_{s}$ and $f$, equations (\ref{b_soft_int})
and (\ref{f}). However these integrals are analytic for selected
power laws $r$. For instance, setting $r=1$ or $r=2$ and then
instructing symbolic math software such as MAPLE to do the radial
integration first and the angular integration second yields
$I_{\tilde{\omega}}=-1/\sqrt{1+\mathfrak{h'}^2/2}\simeq-1$ and 
$I_{\Omega}=[\mathfrak{h'}^{-1}+\sqrt{1+\mathfrak{h'}^2/4}+
  \mathfrak{h'}/2]/\sqrt{(1+\mathfrak{h'}^2/4)
  (1+\mathfrak{h'}^2/2)}\simeq1/\mathfrak{h'}$ for small
$\mathfrak{h'}$. The Lagrange planetary equations then give the
test--ring's precession rates:
\begin{subequations}
\label{precession}
\begin{align}
\label{w-dot}
\dot{\tilde{\omega}}\simeq&
  \frac{\partial R/\partial e}{na^2e}=I_{\tilde{\omega}}\mu_dn
  \simeq-\mu_dn\\
\dot{\Omega}\simeq&
  \frac{\partial R/\partial i}{na^2i}=-I_{\Omega}\mu_dn
  \simeq-\mu_dn/\mathfrak{h'}.
\end{align}
\end{subequations}
Very similar precession rates were previously derived in
\cite{H80} and \cite{W81}. Note that the model rings start to
overlap when their fractional radial thickness
$2\mathfrak{h'}$ exceeds the rings' fractional separation
$\delta=\Delta a/a$, so a massless test ring should precess at
the above rates when the disk--rings are sufficiently
overlapping.

These expectations are tested by constructing a 50
M$_\oplus$ disk having an $r=2$
power--law surface density using 200 circular, coplanar rings
arranged over $10<a<100$ AU, with a number of thin, massless
`test--rings' also orbiting within this disk.
Several simulations are
performed with the massive rings having a variety of thicknesses
$\mathfrak{h'}$. As expected, those test--rings that reside
far from the disk's edges precess
at rates given by equation (\ref{precession}) whenever the
disk--rings are sufficiently overlapping, namely, when
$\mathfrak{h'}\ge2\delta$. 

\subsubsection{precession in an eccentric disk}
\label{eccentric disk}

Although a particle's periapse will experience retrograde
precession when embedded in an axisymmetric disk, prograde
precession is possible in a non--axisymmetric disk.
For instance, prograde precession is evident in the
N--body simulations of an
eccentric stellar disk orbiting the putative black hole at the
center of the Andromeda galaxy M31 \citep{JS01}. These simulated
disks have masses $\sim0.1$ times the central mass, with
the interior parts of the disk being progressively more
eccentric. These simulations reveal a long--lived
overdense region in the direction of apoapse that persists
due to a coherent alignment of the particles' periapse, with
the density pattern rotating
in a prograde sense at rates that increase with the disk mass.
It should be noted that the disk particles' eccentricities are
not always small, so these N--body simulations cannot be
used as a quantitative benchmark for the rings code.
Nonetheless, it is comforting to find that the rings code does
indeed reproduce the density patterns seen in the N--body disks
that precess at rates very similar to that reported in
\cite{JS01}. 

\section{Spiral Wave Theory}
\label{waves}

Section \ref{results} will use the preceding rings model
to demonstrate that apsidal density waves and nodal bending waves
may have once propagated throughout the Kuiper Belt. An apsidal
wave is a one--armed spiral density wave that
slowly rotates over a periapse precession timescale. Similarly,
a nodal wave is a one--armed spiral bending wave that rotates
over a nodal precession timescale.

A brief review of spiral wave theory is in
order. Many of the waves' properties, such as their wavelength
and propagation speed, 
are readily extracted from the waves' dispersion relations.
These dispersion relations are usually obtained from solutions to
the Poisson and Euler equations for the disk. However the
following will show that these dispersion relations may also be
derived from the Lagrange planetary equations.

\subsection{Apsidal Density Waves}
\label{density waves}

The disturbing function $R(a)$ for the disk material
orbiting at a semimajor axis $a$ is
obtained from equation (\ref{R_soft}) with the perturbing mass
$m'$ replaced by the differential mass $dm'$ whose contributions
are integrated across a semi--infinite disk:
\begin{equation}
\label{Rdisk}
R(a)=\frac{1}{2}\mu_dn^2a^2\int_0^\infty
\alpha^{1-r}\left[\frac{1}{2}e^2
f(\alpha,\mathfrak{h},\mathfrak{h})+
ee'\cos(\tilde{\omega}'-\tilde{\omega})
g(\alpha,\mathfrak{h},\mathfrak{h})\right]d\alpha.
\end{equation}
where $e'(\alpha)$ and $\tilde{\omega}'(\alpha)$ are the orbit
elements of the perturbing parts of the disk,
the unprimed quantities refer to the perturbed annulus
at $a$, $r$ is the power--law for the disk's surface density
variation, and a constant fractional thickness $\mathfrak{h}$
is assumed throughout the disk. The Lagrange planetary equations
then give the disk's periapse precession rate at $a$:
\begin{equation}
\dot{\tilde{\omega}}(a)\simeq\frac{\partial R/\partial e}{na^2e}
=\mu_dn(I_{\tilde{\omega}}+I_{dw})
\end{equation}
where $I_{\tilde{\omega}}\simeq-1$ is from the left term in
equation (\ref{Rdisk}) and is the contribution by an
undisturbed disk to its own precession [see equations
(\ref{II}--\ref{precession})], and the right term is
\begin{equation}
\label{I_wave1}
I_{dw}=\frac{1}{2}\int_0^\infty
\frac{e'}{e}\cos(\tilde{\omega}'-\tilde{\omega})
\alpha^{1-r}g(\alpha,\mathfrak{h},\mathfrak{h})d\alpha,
\end{equation}
which is the relative rate at which the density 
wave drives its own precession.

It is expected that the eccentricities associated with a spiral
density wave will vary only slowly with distance $a$ such that
$e'(\alpha)/e\simeq1$. The spiral wave will also organize the
disk's longitude of periapse $\tilde{\omega}$ such that it
varies as
$\tilde{\omega}(a,t)=\dot{\tilde{\omega}}t-\int^ak(A)dA$, where
$k(a)$ is the wavenumber and $\lambda=2\pi/|k|$ is the radial
wavelength. If the spiral wave pattern is tightly wound such that
$\lambda\ll a$ and $|ka|\gg1$, then the dominant contributions to
$I_{dw}$ are largely due to the nearby parts of the disk
where $\alpha=a'/a\sim1\pm\lambda/a$ and
$\tilde{\omega}'-\tilde{\omega}=-\int_{a}^{a'}k(A)dA\simeq
-k(a'-a)$, while the contributions
from the more distant parts of the disk tend to cancel owing to
the rapid oscillation of the cosine factor. Thus we can set
$\alpha=1$ in equation (\ref{I_wave1}) except where it appears
as the combination $x=\alpha-1$ where $|x|\ll1$.
In this case the
softened Laplace coefficients that are present in the
$g$ function can be replaced with the approximate forms
that are valid for $|x|\ll1$ [see equation (\ref{b_local})], so
$g$ becomes
\begin{equation}
g(x)\simeq\frac{2}{\pi}
\frac{2\mathfrak{h}^2-x^2}{(2\mathfrak{h}^2+x^2)^2}.
\end{equation}
It is also permitted to extend the lower integration limit in
equation (\ref{I_wave1}) to $-\infty$ in the tight--winding
limit, so
\begin{equation}
I_{dw}\simeq\frac{1}{2}\int_{-\infty}^\infty
\cos(|ka|x)g(x)dx=|ka|e^{-\sqrt{2}\mathfrak{h}|ka|}.
\end{equation}
And finally, if the wave is to remain coherent across this disk,
this self--precession must occur at the same constant rate
$\omega$ throughout the disk, so
$\dot{\tilde{\omega}}(a)=\omega\simeq
\mu_d(|ka|e^{-\sqrt{2}\mathfrak{h}|ka|}-1)n$.
Note that this disturbing frequency $\omega$, which is also
called the pattern speed, can also be identified as any
one of the eigenfrequencies $g_i$ that appear in equations
(\ref{solution}). Usually it is another perturber that is
responsible for launching the wave and causing the disk to
precess in concert at the rate $\omega$, and this is usually at a
rate that dominates over the non--wave contribution to
the disk's precession, {\it i.e.}, $|\omega|\gg \mu_dn$.
This then yields the
dispersion relation for tightly--wound apsidal density waves:
\begin{equation}
\label{edispersion}
\omega\simeq e^{-\sqrt{2}\mathfrak{h}|ka|}\mu_d|ka|n
\end{equation}
which has the dimensionless form
\begin{equation}
\label{w_e}
\omega_e(K)=Ke^{-K}
\end{equation}
where
$\omega_e=\sqrt{2}\mathfrak{h}\omega/\mu_dn$ is the
dimensionless frequency and $K=\sqrt{2}\mathfrak{h}|ka|$ the
dimensionless wavelength; Figure \ref{disp_rel} plots
$\omega_e$ versus $K$. A more general dispersion relation for
an $m$--armed spiral wave
in a stellar disk is given in \cite{T69}, and in the limit that
$|\omega|\ll n$ the resulting formula with $m=1$
behaves qualitatively quite similar\footnote{In this limit,
Toomre's dispersion relation becomes $\omega_T(K)=K{\cal F}$
where the reduction factor ${\cal F}$ is a more complicated
function of $K$. However a numerical evaluation of this function
shows that $\omega_T(K)$ has the same form as the $\omega_e(K)$
curve shown in Figure \ref{disp_rel}.} to equation (\ref{w_e}).
Note that equation (\ref{edispersion}) also
recovers the usual dispersion relation $\omega=\mu_d|ka|n$ for
apsidal waves in an infinitesimally thin disk when
$\mathfrak{h}=0$.

Note that $\omega_e(K)>0$ and that it 
also has a maximum at $K=1$ where
$\omega_e(1)=\exp(-1)$. Since $\omega_e$
is a function of semimajor axis $a$, the restriction
$0<\omega_e(a)\lesssim0.368$ indicates that apsidal waves can
only propagate in a restricted interval in $a$. This
restriction can also be viewed as a constraint on the disk
thickness, namely,
\begin{equation}
\label{hQ}
\mathfrak{h}\lesssim0.260\mu_d|n/\omega|
\equiv\mathfrak{h}_Q.
\end{equation}
Alternatively, equation (\ref{hQ}) can also be viewed as an
upper limit on the frequency $\omega$ or a lower limit on the
disk mass $\mu_d$ wherein wave--action is permitted.

Waves having a wavenumber $K<1$ are called long waves
since they have a wavenumber $|k_L|\simeq(\omega/n)/\mu_da$ and
the longer wavelength
$\lambda_L=2\pi/|k_L|\simeq2\pi\mu_d(n/\omega)a$, while short
waves have wavenumbers $K>1$ or $|k_S|>1/\sqrt{2}\mathfrak{h}a$
and the shorter wavelength
$\lambda_S=2\pi/|k_S|<2\sqrt{2}\pi\mathfrak{h}a$. These long and
short density waves also correspond to the $g$ and $p$
modes, respectively, of \cite{T01}. Writing $\lambda_L$ in term
of $\mathfrak{h}_Q$ and then requiring
$\mathfrak{h}<\mathfrak{h}_Q$ also means that apsidal wave can
propagate wherever $\lambda_L\gtrsim24h$.

The rate at which apsidal waves propagate across the disk is
given by their group velocity
\begin{equation}
\label{density c_g}
c_g=\frac{d\omega}{dk}=s_k\mu_d(1-K)e^{-K}na
\end{equation}
where $s_k=sgn(k)$. Waves with $s_k=+1$ are called
trailing waves, and their longitude of perihelia $\tilde{\omega}$
decreases with increasing semimajor axis $a$, while $s_k=-1$ are
leading waves whose longitude increases with $a$.
Since the waves' group velocity $c_g$ is proportional to
the slope of the $\omega_e(K)$ curve, the site where $K=1$
is a turning point where wave reflection occurs;
in galactic dynamics this reflection site is known as a $Q$
barrier. A long wave that approaches the $Q$ barrier
from the $K<1$ side of Figure \ref{disp_rel} thus reflects
as a short wave as it continues along the $K>1$ side of the
curve. 
The simulations shown in Section \ref{results} also show
that long trailing waves that instead strike a disk edge
also reflect as short trailing waves.

\subsection{surface density variations}
\label{surface density}

The compression or rarefaction among the disk's rings, or
streamlines, is $(\partial r/\partial a)^{-1}$, which
is also be the relative change in the disk's surface density
$\sigma/\sigma_o$ associated with a density wave
[c.f., \cite{BGT85}]. For a small amount of compression,
$\sigma/\sigma_o=1+\Delta\sigma/\sigma_o$ where the
fractional surface density variation is
\begin{equation}
\label{Delta-sigma}
\frac{\Delta\sigma}{\sigma_o}=
\left(\frac{\partial r}{\partial a}\right)^{-1}-1
\simeq\frac{\partial(ea)}{\partial a}\cos(\phi-\tilde{\omega})
+ea\frac{\partial\tilde{\omega}}{\partial a}
\sin(\phi-\tilde{\omega})
\end{equation}
to lowest order in $e$. The second term dominates over
the first in the tight--winding limit so
$|\Delta\sigma/\sigma_o|\sim\mathcal{O}(e|ka|)$. Density waves
are nonlinear when $|\Delta\sigma/\sigma_o|>1$, and these large
density variations are a consequence of overlapping streamlines.
For long density waves having a wavenumber
$|k_L|=\omega/\mu_dan$, the disk's streamlines will cross when
the waves' eccentricities exceed $e_L\sim\mu_d\omega/n$, while
streamline crossing occurs among short waves when eccentricities
exceed $e_S\sim\sqrt{2}\mathfrak{h}$. As the
simulations of Section \ref{results} will show, a dynamically
cool Kuiper Belt is very susceptible to the propagation of
short nonlinear density waves that facilitate streamline
crossing. Depending upon the relative velocities of
these crossed streamlines, apsidal wave--action might encourage
accretion or else enhance collisional erosion among KBOs.

\subsection{Nodal Bending Waves}
\label{bending waves}

The derivation of the dispersion relation for nodal bending
waves, and its analysis, proceeds similarly. The disk's
integrated disturbing function is
\begin{equation}
\label{Ri}
R(a)=-\frac{1}{2}\mu_dn^2a^2\int_0^\infty\alpha^{2-r}
\tilde{b}^{(1)}_{3/2}(\alpha,\mathfrak{h},\mathfrak{h})
\left[\frac{1}{2}i^2-ii'\cos(\Omega'-\Omega)
\right]d\alpha.
\end{equation}
so the Lagrange planetary equation gives
\begin{equation}
\dot{\Omega}(a)\simeq\frac{\partial R/\partial i}{na^2i}
=\mu_dn(I_{bw}-I_{\Omega})
\end{equation}
where $\Omega(a,t)=\dot{\Omega}t-\int^ak(A)dA$ and
\begin{equation}
\label{I_bw}
I_{bw}=\frac{1}{2}\int_0^\infty \alpha^{2-r}
\tilde{b}^{(1)}_{3/2}(\alpha,\mathfrak{h},\mathfrak{h})
\cos[ka(\alpha-1)]d\alpha
\simeq\frac{e^{-\sqrt{2}\mathfrak{h}|ka|}}{\sqrt{2}\mathfrak{h}}
\end{equation}
is the bending wave's contribution to its own precession.
The contribution from the undisturbed disk is
$I_{\Omega}\simeq1/\sqrt{2}\mathfrak{h}$, with
the additional $\sqrt{2}$ factor the result of
changing the middle argument in equation (\ref{IOmega}) from 0
to $\mathfrak{h}$. Since $\omega=\dot{\Omega}$ is a constant
for a coherent wave, the dispersion relation for tightly--wound
nodal bending waves is
\begin{equation}
\label{dispersion}
\omega\simeq\frac{\mu_d}{\sqrt{2}\mathfrak{h}}
  \left( e^{-\sqrt{2}\mathfrak{h}|ka|}-1\right)n.
\end{equation}
Note that the usual dispersion relation for nodal waves in an
infinitesimally thin disk, $\omega\simeq-\mu_d|ka|n$, is
obtained when $\mathfrak{h}\rightarrow0$.

The dimensionless form of the dispersion relation is 
\begin{equation}
\label{w_i}
\omega_i(K)=e^{-K}-1,
\end{equation}
where $\omega_i=\sqrt{2}\mathfrak{h}\omega/\mu_dn$,
which is plotted in Figure \ref{disp_rel}. As the Figure
shows, nodal bending waves can propagate only in regions where
$-1<\omega_i(a)<0$ which similarly limits the disk
thickness to $\mathfrak{h}\lesssim2.72\mathfrak{h}_Q$.
The nodal waves' group velocity is
\begin{equation}
c_g=\frac{d\omega}{dk}=-s_k\mu_d(1+\omega_i)na,
\end{equation}
which indicates that nodal waves tend to stall, {\it i.e.},
$|c_g|\rightarrow0$ as they approach the $\omega_i=-1$ boundary.
Note that this dispersion relation only admits a long wavelength
solution having a wavenumber $|k_L|\simeq-\omega/\mu_dna$ and
a wavelength $\lambda_L\simeq2\pi\mu_d|n/\omega|a$ for
waves far from the stall--zone. Since
$\mathfrak{h}\lesssim2.72\mathfrak{h}_Q$, the disk can sustain
nodal waves wherever $\lambda_L\gtrsim9h$. But if an
outward--traveling long leading wave with $s_k=-1$ 
encounters a disk edge, it will reflect as an $s_k=+1$
long trailing wave. Examples of nodal wave reflection and
stalling are also given below.

\section{The Secular Evolution of the Primordial Kuiper Belt}
\label{results}

Using the recipe given in Section \ref{N rings}, the rings model
is used to compute the secular evolution of the primordial
Kuiper Belt as it is perturbed by the four giant planets.
In these simulations the giant planets are represented by
thin $\mathfrak{h}=0$ rings whose initial orbits are their
current orbits, while the Kuiper Belt is represented by 500 rings
whose semimajor axes extend from 36 AU out to 50--70 AU.
The location of the Belt's inner edge is chosen
such that only the outermost radial and vertical secular
resonances, the $\nu_8$ and the $\nu_{18}$, reside in
this disk near 40 AU when of low mass.
The semimajor axes of each Belt--ring increases as
$a_{j+1}=(1+\delta)a_j$ where the rings'
fractional separation $\delta$ is typically $\sim0.001$.
The rings fractional half--thickness $\mathfrak{h}$ is always in
excess of $2\delta$, as is required to get the
correct apse precession in an axisymmetric disk (see Section
\ref{axisymmetric disk}).
The Belt--rings initial eccentricities and inclinations are
zero, with all inclinations being measured with respect to the
system's invariable plane. The mass of each ring is chosen such
that the Belt's surface density $\sigma(a)$ varies as $a^{-1.5}$.
For this configuration, if
the total Kuiper Belt mass over the $36\le a\le 70$ AU zone
is $M_{\mbox{\scriptsize total}}$, then the total mass in the
`observable' $30\le a\le 50$ AU zone that is presently
accessible to astronomers would be
$M_{KB}=0.67M_{\mbox{\scriptsize total}}$ had the above
surface density law extended inwards to 30 AU. For these systems
the normalized disk mass is $\mu_d\simeq M_{KB}/M_\odot$.

\subsection{A $M_{KB}=10$ M$_\oplus$ Example}
\label{example}

Figure \ref{M_KB=10} shows a snapshot of apsidal density
waves as they propagate across an $M_{KB}=10$
M$_\oplus$ Kuiper Belt having a half--thickness
$\mathfrak{h}=5\delta=0.0067$. Since
$\mathfrak{h}\sim0.2\mathfrak{h}_Q$, the necessary
disk--conditions for the propagation of apsidal and nodal waves
are well--satisfied. Initially, a long trailing density wave is
launched at the Belt's inner edge.
This wave is really more like a pulse $\sim5$ AU wide, and Figure
\ref{M_KB=10} shows that by time $t=2\times10^6$ years the
wave has just started to reflect at the disk's outer edge at 70
AU. The greyscale map of the disk's surface density variation,
$\Delta\sigma/\sigma$, is obtained using equation
(\ref{Delta-sigma}), and this map also provides a historical
record of this system's wave--action. It should be noted that
equation (\ref{Delta-sigma}) is quantitatively correct only when
$|\Delta\sigma/\sigma|\ll1$, a condition that is rarely
satisfied by the results obtained here.
Nonetheless, equation (\ref{Delta-sigma}) is still useful in a
qualitative sense since it will indicate the disk--regions where
large surface density variations as well as orbit--crossing
can be expected. As the outer edge of the
$\Delta\sigma/\sigma$ map shows, the main apsidal
density wave--pulse at $67<a<70$ AU has
just reflected at a short trailing wave, and this
nonlinear wave having $|\Delta\sigma/\sigma|>1$ will completely
dominate the disk's appearance at later times as it propagates
inwards. But until this happens, the bulk of the disk's
appearance  over $45<a<67$ AU is still dominated by
lower--amplitude long waves that are following behind the main
density pulse. Note also that short leading waves were also
emitted at the disk's inner edge, but as the density greyscale
shows, they have only propagated out to $a=45$ AU thus far owing
to their slower group velocity
[see equation (\ref{density c_g})]. These short waves are
well--resolved in the sense that their radial wavelength of
$\lambda_S\sim1$ AU spans about 15 disk--rings.
Although these short waves are seen in the $e(a)$ and
$\tilde{\omega}(a)$ plots as only tiny wiggles over
$36<a<45$ AU that are superimposed on top
of the disk's longer--wavelength behavior, the density greyscale
shows that the short waves dominate the inner disk's
appearance. The online edition of Fig \ref{M_KB=10} is also
linked to an animated sequence of these figures that give the
system's complete time--history. That animation shows that by
time $t\simeq1\times10^7$ years, nonlinear short waves will have
swept across the entire disk, and they result in large surface
density variations $|\Delta\sigma/\sigma|>1$ over radial
wavelengths of $\lambda_S\sim1$ AU.

Figure \ref{M_KB=10} also shows that by time
$t=2\times10^6$ years, a long leading nodal bending
wave--pulse has already propagated across the disk where it has
reflected at the disk's outer edge and just started to return
inwards as a long trailing bending wave. But when that pulse
reaches the disk's inner edge, a portion of the wave's
angular momentum content will continue to propagate further
inwards where it will give a small kick to the giant planets'
orbit elements, while the remaining wave--pulse reflects again
and propagates outwards. The same phenomenon also occurs among
the apsidal density waves. Thus after a few reflections, a single
wave--pulse will lose its initial spatial
coherence by spawning multiple
wave--trains that, in this friction--free model, forever roam
about the Belt. This ultimately results in a rather
wobbly--looking standing density wave pattern
that varies over the short wavelength scale $\lambda_S\sim1$ AU,
as well as a standing bending wave pattern
that varies over a much longer scale $\lambda_L\sim40$ AU.

A dynamical `spectrum' of this Kuiper Belt is shown in
Figure \ref{eigenvectors}, which plots all of the
eccentricity eigenvector elements $|E_{ji}|$ for each of the
disk--rings versus their eigenfrequency $g_i$, as well as
inclination eigenvector elements $|I_{ji}|$ versus
eigenfrequency $|f_i|$. The upper figure is quite reminiscent of
the findings of \cite{T01}, who showed that a gravitating
disk tends to exhibit its strongest response to slow radial
perturbations via modes having discrete patterns speeds
$\omega$ that can be identified with any of the peak frequencies
$g_i$ seen in Figure \ref{eigenvectors}.
Figures such as these are quite useful for
identifying the patterns speeds $\omega$ that are associated
with the density and bending waves that propagate in the disk.

\subsection{Variations with Kuiper Belt Mass $M_{KB}$}
\label{mass-variations}

Simulations have been performed with
Kuiper Belts having masses $M_{KB}=0$, 0.08, 0.2, 2, 10, and 30
M$_\oplus$ in the observable $30\le a\le50$ AU zone (again
this would be the Belt's mass assuming its
surface density $\sigma(a)\propto a^{-1.5}$
were to extend solely over that region). The results are
summarized in Figure \ref{compare_ei} which shows the maximum
eccentricities $e_{\mbox{\scriptsize max}}$ and maximum
inclinations $i_{\mbox{\scriptsize max}}$ achieved by the
rings in each simulation. The Belt's radial width is indicated by
the breadth of the curves in Figure \ref{compare_ei}, which
ranges from 34 AU in the higher--mass Belts to 14 AU for the
$M_{KB}=0.08$ M$_\oplus$  system. Each simulation uses 500
disk--rings having a
fractional half--thickness $\mathfrak{h}=2\delta$, so the
three higher--mass systems have $\mathfrak{h}=0.0027$ while the
lower--mass disk $M_{KB}=0.2$ M$_\oplus$ is somewhat thinner with
$\mathfrak{h}=0.0015$ and the $M_{KB}=0.08$ M$_\oplus$ 
system has $\mathfrak{h}=0.0010$. The rings in these
lower--mass disks are more closely packed so that their
shorter--wavelength density waves are well resolved,
and they are also made thinner so as to push their
$Q$--barriers a bit further downstream. The
disk's initial orbits are $e=0=i$, except for the $M_{KB}=0$
system which instead adopts the forced orbit elements
appropriate for a massless disk
[e.g., \cite{BC61, MD99}]. These systems are
evolved until their initial density and bending wave pulses have
had the opportunity to reflect multiple times and have largely
dissolved into standing wave patterns. The lower--mass disks
necessarily have longer run--times due to their waves' slower
propagation speeds [see equation (\ref{density c_g})];
these run--times are
$1\times10^9$ years,  $2\times10^9$ years, $1\times10^8$ years,
$2\times10^7$ years, and $1\times10^7$ years, respectively, for
the $M_{KB}=0.08$, 0.2, 2, 10, and 30 M$_\oplus$ systems. 
Once the standing wave pattern has emerged, the rings'
instantaneous eccentricities and inclinations range over
$0\lesssim e<e_{\mbox{\scriptsize max}}$ and
$0\lesssim i<i_{\mbox{\scriptsize max}}$.

As Figure \ref{compare_ei} shows, there are no peaks in the
$e_{\mbox{\scriptsize max}}$ and $i_{\mbox{\scriptsize max}}$
curves for the higher mass disks having $M_{KB}\ge2$ M$_\oplus$,
indicating that there are no secular resonances in the disk
itself; any such resonances likely lie between the orbits of
Neptune and the disk's inner edge at 36 AU. The
simulations of these higher--mass disks
show long trailing density waves and long leading bending waves 
being launched at the disk's inner edge. These waves sweep
across the disk, reflect at the disk's outer edge, and return 
as short trailing density waves and long trailing bending waves,
similar to the history described in Section \ref{example} and
seen in Figure \ref{M_KB=10}. 

However secular resonances at $a\simeq41$ AU are quite prominent
in the lower--mass disks having $M_{KB}=0.08$ and 0.2 M$_\oplus$.
These are sites that launch long density and
bending waves, and as Figure \ref{M_KB=10} shows, the density
waves are able to propagate out as far as $a=45$ and 49 AU,
respectively, where they are reflected by a $Q$--barrier and
return as short density waves. These reflection sites occur
where $\mathfrak{h}\simeq2.2\mathfrak{h}_Q$ and 
$\mathfrak{h}\simeq1.4\mathfrak{h}_Q$, which shows that equation 
(\ref{hQ}) is an approximate yet useful indicator of where
apsidal wave are allowed to propagate. Note also the large
amplitudes of the density waves in these low--mass disks,
$0.3\lesssim e_{\mbox{\scriptsize max}}\lesssim1$, which clearly
violates the model's assumption of low eccentricities. Thus these
particular curves should not be taken at face value.
Nonetheless, they do indicate that apsidal waves in a
low--mass Kuiper Belt may result in large eccentricities, and
Figure \ref{M_KB=10} shows that sizable inclinations can also
result due to nodal bending waves. Indeed, in the
$M_{KB}=0.08$ M$_\oplus$
disk, maximum inclinations are typically $i\sim20^\circ$, which
is comparable to the Main Belt's high--inclination component.

The larger $e$'s and $i$'s seen in the lower--mass disk's are a
consequence of the giant planets transmitting a small fraction of
their initial angular momentum deficit\footnote{e.g., the
$L_e$ and $L_i$ of equations (\ref{Lei}).} (AMD) into the disk in
each simulation. In each simulation, the giant planets deposit
$\sim0.005L_e$ and $\sim0.1L_i$ into the disk's inner edge,
which waves then transport and smear out across a vast swath of
the Kuiper Belt. Since the AMD deposited in the disk is roughly
constant in each simulation, the lower--mass disks exhibit
larger $e$ and $i$ excitations.
It should also be noted that computational
limitations in dynamical studies of the Kuiper Belt,
particularly N--body simulations, are often require treating
the Belt as a swarm of massless test particles. However the
comparison of the $M_{KB}=0$ curve to the
$M_{KB}>0$ curves in Figure \ref{M_KB=10} shows that the
endstate of a Kuiper Belt having even just a modest amount of
mass can be radically different from one
naively treated as massless.

\subsection{Variations with Disk Thickness $\mathfrak{h}$}
\label{thickness-variations}

Figure \ref{M=10} shows how the response of an
$M_{KB}=10$ M$_\oplus$ Belt varies with increasing disk thickness
$\mathfrak{h}=(2, 20, 30, 60, 100)\times\delta=$(0.0027, 0.027,
0.04, 0.08, 0.13). According to equation (\ref{hQ}),
$\mathfrak{h}_Q(a)\propto a^{1/2-r=-1}$ in these $r=3/2$ disks,
so the $Q$--barrier will move inwards as the disk's
$\mathfrak{h}$ is increased, as is evident in  Figure \ref{M=10}.
All of these disks have a normalized disk mass
$\mu_d\simeq3\times10^{-5}$, a mean motion $n\simeq0.02$
radians/year, and a pattern speed that is typically
$\omega\sim3\times10^{-6}$ radians/year, so wave--action is shut
off when the disk thickness $\mathfrak{h}$ exceeds
$\mathfrak{h}_Q=0.26\mu_d|n/\omega|\sim0.05$. 
In the thinnest disk with
$\mathfrak{h}=0.0027$, the $Q$--barrier lies beyond the disk's
outer edge at 70 AU, so long and then short apsidal density
waves are able to slosh about the disk's full extent. However the
$Q$--barrier lies in the disk at $a\simeq60$ AU when 
$\mathfrak{h}=0.027$ (red curve), at $a\simeq53$ AU when
$\mathfrak{h}=0.04$ (green curve), and apsidal waves are
prohibited in the disks with $\mathfrak{h}\ge0.08$
(blue and purple curves).

Figure \ref{M=10} also shows that the nodal bending waves are
shut off when the disk thickness $\mathfrak{h}$ exceeds the
somewhat more relaxed criterion $2.72\mathfrak{h}_Q\sim0.14$.
Note also the peak at $a\simeq53$ AU in the
$\mathfrak{h}=0.027$ disk (red curve) and at $a\simeq39$ AU in
the $\mathfrak{h}=0.04$ disk (green curve). These particular
disks admit two spiral patterns, a higher--amplitude spiral
that corotates with Neptune's dominant eigenmode at the
pattern speed of
$\omega\sim-3\times10^{-6}$ radians/year, and a lower--amplitude
mode that corotates with Uranus at the faster rate
$\omega\sim-1.5\times10^{-5}$ radians/year. Since
$\mathfrak{h}_Q\propto|\omega|^{-1}$, the faster
spiral pattern has $2.72\mathfrak{h}_Q\sim0.03$, which is why
this wave stalls at $a=53(39)$ AU in the
$\mathfrak{h}=0.027(0.04)$ disks while the slower spiral mode is
able to propagate the full breadth of these disks.

The behavior of a lower--mass disk with $M_{KB}=0.2$ M$_\oplus$
is shown in Figure \ref{M=0.2} for disks having
$\mathfrak{h}=(2, 5, 10, 15)\times\delta=$(0.0015, 0.0037,
0.0074, 0.011). A pair of secular resonances lie near
$a\simeq40$ AU, and they launch apsidal and nodal waves
having pattern speeds $\omega\sim\pm3\times10^{-6}$
radians/year. These disks have $\mu_d\simeq6\times10^{-7}$
and $\mathfrak{h}_Q\sim0.001$, and as the upper plot shows,
further increases in $\mathfrak{h}$ brings the $Q$--barrier
inwards until the wave--propagation zone has shrunk down to
zero. The lower plot also shows that nodal waves forever
slosh about the model Kuiper Belt in the $\mathfrak{h}=0.0015$
disk (orange curve), whereas the peaks in the other curves
show that these waves stall at sites ever closer to resonance
in progressively thicker disks.
These Figures also show that when
$\mathfrak{h}\gg\mathfrak{h}_Q$, the motions of a
non--gravitating disk are recovered, namely,
the disk's $e$'s and $i$'s are
twice the forced motions seen in the $M_{KB}=0$ disk (black
curves), with the factor of two being a consequence of these
disks' initial conditions $e=0=i$.

\subsection{Implications for the Primordial Kuiper Belt}
\label{implications}

The primordial Kuiper Belt likely had
an initial mass of $M_{KB}\sim30$ M$_\oplus$ (see Section
\ref{introduction}), and accretion models show that the initial
KBO swarm must have had dispersion velocities less than
$\sim0.001\times$ Keplerian \citep{KL99}, so 
$\mathfrak{h}\lesssim0.001$, $\mu_d\sim1\times10^{-4}$, and thus
$\mathfrak{h}_Q\sim0.2$ assuming the spiral waves have pattern
speeds of  $|\omega|\sim3\times10^{-6}$ radians/year. Since
$\mathfrak{h}<\mathfrak{h}_Q$, the primordial Kuiper Belt readily
sustained apsidal and nodal waves. Figure \ref{compare_ei} shows
that in this high--mass environment, these will be rather
low--amplitude waves having $e\sim\sin i\sim0.01$. These
waves will quickly propagate across a Kuiper Belt of width
$\Delta a$ in time $T_{\mbox{\scriptsize prop}}
\sim\Delta a/|c_g|\simeq\Delta a/\mu_dan$, so the wave
propagation time is
\begin{equation}
T_{\mbox{\scriptsize prop}}
\sim3\times10^5\left(\frac{M_{KB}}{30 M_\oplus}\right)^{-1}
\left(\frac{\Delta a}{30\mbox{ AU}}\right)\mbox{ years}.
\end{equation}
In the friction--free model employed here, the outward--bound
long density waves eventually reflect, either at a $Q$--barrier
in the disk (which might lie downstream where
$\mathfrak{h}=\mathfrak{h}_Q$) or else at the disk's outer edge.
The reflected waves then propagate inwards as short density
waves, and such waves are nonlinear in the sense that their
surface density variations $\Delta\sigma/\sigma$ typically exceed
unity. Figure \ref{M_KB=10} shows a snapshot of long and short
density waves in a $M_{KB}=10$ M$_\oplus$ disk.
Note that the long waves completely dominate
the disk's orbit elements $e(a)$ and $\tilde{\omega}(a)$
that vary over a wavelength of $\lambda_L\sim10$ AU, while the
short waves are the tiny variations in $e(a)$ and
$\tilde{\omega}(a)$ that occur over a $\lambda_S\sim1$ AU scale
at the disk's inner and outer edge. Even though the short waves
are almost imperceptible in the orbit--elements plots, the
greyscale map shows that
these nonlinear short waves completely dominate the disk's
surface density structure.

Figure \ref{compare_ei} also shows that the waves in lower--mass
systems have higher amplitudes. This suggests that
wave action may tend to drive disk--planet systems towards an
equipartition of angular momentum deficits, since the angular
momentum content of the waves seen in all simulations are
$\sim0.5\%$ and $\sim10\%$ of the planets' $L_e$ and
$L_i$.

The large--amplitude waves seen in the $M_{KB}=0.2$ M$_\oplus$
disk also suggests that apsidal and
nodal wave action might account for much of the Kuiper
Belt's excited state (see Figure \ref{compare_ei}). However the
viability of this scenario depends critically upon the
timescales that govern the Belt's mass loss as well as the
KBOs' velocity evolution. First note that
accretion models show that as soon as the large
$R\gtrsim100$ km KBOs form, further KBO growth is halted as they
initiate an episode of more vigorous collisions that steadily
grind the Belt's smaller members down to dust which radiation
forces then transport away \citep{KL99, KB01}. Accretion models
show that the $R\sim100$ km KBOs form at $a\sim35$ AU over a
$\tau_{\mbox{\scriptsize form}}\sim1\times10^7$ year timescale,
and that the Belt's subsequent collisional erosion occurs over a
$\tau_{\mbox{\scriptsize erode}}\sim5\times10^8$ year timescale
\citep{KL99, KB01}. And if Neptune's orbit had migrated
substantially, the advancing 2:1 resonance would also have swept
across the Main Belt, which further enhances the stirring
as well the collisional/dynamical erosion.
The resulting grinding and erosion thus makes it
ever more difficult for the Belt to sustain waves
as the initially massive Kuiper Belt mass is
eroded by a factor of $\sim150$, which also
lowers the disk's $\mathfrak{h}_Q\rightarrow\sim10^{-3}$. Note
that the removal of the smaller KBOs also turns off the dynamical
friction that was once keeping the particle disk quite thin. The
gravitational stirring by the surviving larger KBOs is then free
to raise their dispersion velocities $c$ up to their surface
escape velocity $v_{\mbox{\scriptsize esc}}$ \citep{S72},
which results in a thicker disk with
$\mathfrak{h}\sim v_{\mbox{\scriptsize esc}}/na
\sim0.02(R/100\mbox{ km})$ where $R$ is the KBOs'
characteristic size. Since this is substantially larger than the
current Belt's $\mathfrak{h}_Q$, it seems quite likely that
this stirring shut off the waves while they were still of
low amplitude (see Fig.\ \ref{compare_ei}) long before the
Kuiper Belt was ground down to its present mass\footnote{An
exception to this assertion might occur if the reflection of
nodal waves at an outer edge allowed the Belt to behave as a
resonant cavity \citep{W03}. When the disk's mass and the
wave's pattern speed are appropriately tuned, which can occur
naturally as the Belt eroded and/or the as Neptune's precession
rate varied due to its orbital migration, then a
higher--amplitude standing wave pattern can result while the Belt
persists in the tuned state.}.
Consequently, apsidal and nodal waves were likely able to
propagate throughout the Kuiper Belt during a timescale
$\tau_{\mbox{\scriptsize wave}}$ that is bounded by the moment
when the large KBOs first formed and when the Belt eroded away,
{\it i.e.}, $\tau_{\mbox{\scriptsize form}}<
\tau_{\mbox{\scriptsize wave}}<\tau_{\mbox{\scriptsize erode}}$.

\subsection{Comments on Studies of Spiral Waves in Galactic Disks}
\label{galaxies}

It should also be noted that this implementation of the rings
model does not account for any viscous damping of spiral waves
which, as \cite{HT69} point out, may be more important as bending
waves approach a disk's outer edge. Unlike the sharp--edged disks
employed here, a more realistic disk likely has a `fuzzy' edge
where the disk's surface density gently tapers to zero.
\cite{HT69} show that bending waves entering such a lower--density
zone tend to excite substantially larger inclinations there, and
such sites will be considerably more susceptible to the viscous
dissipation of these possibly nonlinear waves. 

\cite{T83} suggests that disks having a tapered edge might also
cause bending waves to stall there since the group velocity
$c_g\rightarrow0$ as the surface density $\sigma$ smoothly goes
to zero. However this assertion was not confirmed by the rings
model, which tapered a disk's outer surface density by
multiplying it by the factor $\sqrt{1-(\ell-\Delta a)^2/\ell^2}$
where $\Delta a$ is the distance from the outer edge and $\ell$ is
the tapering scale--length. These experiments adopt values of
$\ell$ that are smaller, comparable, and larger than the bending
wavelength $\lambda_L$, and in all cases the bending wave
reflects at or near the disk edge, with considerably larger
inclinations being excited in this tapered zone.

\section{Summary}
\label{summary}

A model that rapidly computes the secular evolution of a
gravitating disk--planet system is developed. The disk is
treated as a nested set of rings, with the rings'/planets'
time--evolution being governed
by the Lagrange planetary equations. It is shown that the
solution to the dynamical equations is a modified version of
the classical Laplace--Lagrange solution for the secular
evolution of the planets \citep{BC61, MD99}, with the
modification being due to a ring's finite thickness $h=c/n$
that is a consequence of the dispersion velocity $c$ of that
ring's constituent particles. Since the ring's finite thickness
$h$ softens its gravitational potential, this also softens
the Laplace coefficients appearing in the Laplace--Lagrange
solution over a scale $h/a$.

It is shown that the Lagrange planetary equations admit spiral
wave solutions when the tight--winding approximation is applied.
There are two types of spiral density (or apsidal)
waves, long waves of wavelength
$\lambda_L=2\pi\mu_d|n/\omega|a$ and short waves of wavelength
$\lambda_S<2\sqrt{2}\pi\mathfrak{h}a$ where $\mathfrak{h}=h/a$
is the disk's fractional thickness
and $\omega$ is the angular rate at which the spiral pattern
rotates. The simulations presented here show that the giant
planets launch long waves at either a resonance in the disk
or else at the disk's nearest edge, and that these waves
propagate away until they reflect at the disk's far edge
or else at a $Q$--barrier in the disk which resides where
$\mathfrak{h}=\mathfrak{h}_Q(a)$ where
$\mathfrak{h}_Q(a)=0.26\mu_d|n/\omega|$ is the maximum disk
thickness that can sustain apsidal waves, with
$\mu_d=\pi\sigma a^2/M_\odot$ being the normalized disk mass.
Of course all of these findings may be derived from the stellar
dispersion relation given in \cite{T69} in the limit that
the pattern speed $\omega$ is much smaller than the disk's
mean motion $n$. Nonetheless, it is satisfying to see that the
theory of unforced apsidal waves is readily obtained from the
Lagrange planetary equations; with a little more effort the
theory for forced apsidal waves [e.g., \cite{WH98aj}]
should also be recoverable.

However new results are obtained for the nodal wave problem,
which admits only a long--wavelength solution $\lambda_L$ to the
planetary equations in the tight--winding limit. In particular,
it is shown that these waves can stall, that is, the waves' group
velocity plummets to zero as they approach a site in the disk
where $\mathfrak{h}=2.72\mathfrak{h}_Q$. If, however, these
waves instead encounter a disk edge, they will reflect and
return as long waves. In the limit that
$\mathfrak{h}\rightarrow0$, the results for
nodal waves propagating in a infinitesimally thin disk is
recovered \citep{WH03}, but note that the
wave--stalling phenomenon does not appear in a
$\mathfrak{h}=0$ treatment of the disk. 

The rings model is also used to examine the propagation of
apsidal and nodal waves that are launched by the giant planets
into a variety of Kuiper Belts having a
mass $M_{KB}=30$ M$_\oplus$ (the estimated primordial mass) down
to $M_{KB}=0.08$ M$_\oplus$ (which is $\sim40\%$ of the Belts
current mass estimate). In each simulation the giant
planets deposit roughly the same fraction of their initial
angular momentum deficits, $\sim0.5\%$ and $\sim10\%$ of the
planets' $L_e$ and $L_i$, into the disk in the form of spiral
waves. And since the waves' angular momentum content is roughly
the same in each simulation, the lower--mass Kuiper Belts thus
experience higher--amplitude waves. Indeed, the waves seen in the
$M_{KB}\le0.2$ M$_\oplus$ simulations are of sufficient
amplitude that they could in principle account for much of the
dynamical excitation that is observed in the Kuiper Belt. However
wave--action in a $M_{KB}\sim0.2$ M$_\oplus$ Belt also
requires its fractional scale--height to quite thin,
namely, $\mathfrak{h}\lesssim10^{-3}$. Most likely, apsidal and
nodal waves were shut off, due to self--stirring by large KBOs
as well as by other external perturbations, long before the Belt
eroded down to its current mass, in which case the excitation by
wave--action would have been quite modest.

The rings model developed here has many other applications. One
issue of great interest is to determine whether apsidal and
nodal waves may be propagating in Saturn's rings. Of particular
interest are the short apsidal waves since their detection
could yield the ring's dispersion velocity
$c$ via a measurement of the short wavelength
$\lambda_S\sim9c/n$. Although the ring
particles' dispersion velocity is of
fundamental importance to ring dynamics, it is less than well
constrained at Saturn. Of course, the differential precession
due to planetary oblateness also needs to be included in the
model [c.f., \cite{MD99}] since this effect may actually defeat
this form of wave--action. The rings model
can also be used to examine the forced motions
of a relatively massless but much thicker
circumstellar dust disk like $\beta$ Pictoris. The warps 
and brightness asymmetries seen in this system are usually
attributed to secular perturbations exerted by an
unseen planetary system,
and the code developed here can be used to very
rapidly explore the wide range of planetary parameters.
This rings model will be used to study these and other problems
in greater detail in the near future.

\acknowledgments

\begin{center}
{\bf ACKNOWLEDGMENTS}
\end{center}

This paper is contribution XXXX from the Lunar and Planetary
Institute, which is operated by the Universities Space Research
Association by cooperative agreement NCC5--679 with the
National Aeronautics and Space Administration. This research was
also supported by NASA via the Origins of Solar Systems grant
No.\ NAG5--10946 issued through the Office of Space Science.

\newpage
\appendix
\section{Appendix \ref{appendixB}}
\label{appendixB}

Differentiating the $\tilde{b}^{(m)}_s$ appearing in the $f$
and $g$ functions, equations (\ref{fg}), yields
\begin{subequations}
\label{fg1}
\begin{align}
f=&2\alpha\tilde{b}^{(1)}_{3/2}+\frac{3}{2}\alpha^2
  \left(\tilde{b}^{(2)}_{5/2}-\tilde{b}^{(0)}_{5/2}\right)
  -3\alpha^2H^2(2+H^2)\tilde{b}^{(0)}_{5/2}\\
g=&2(\alpha^2+1)(1+H^2)\tilde{b}^{(1)}_{3/2}
 -3\alpha\left(\tilde{b}^{(0)}_{3/2}+\tilde{b}^{(2)}_{3/2}\right)
 +3\alpha^2H^2(2+H^2)\tilde{b}^{(1)}_{5/2}
 -\frac{3}{4}\alpha^2
   \left(\tilde{b}^{(3)}_{5/2}-\tilde{b}^{(1)}_{5/2}\right)
\end{align}
\end{subequations}
where $H^2=\onehalf(\mathfrak{h}^2+\mathfrak{h'}^2)$.
The following recursion relations,
\begin{subequations}
\label{recursion1}
\begin{align}
\label{recursion1a}
m\tilde{b}^{(m)}_s=&s\alpha
  \left(\tilde{b}^{(m-1)}_{s+1}-\tilde{b}^{(m+1)}_{s+1}\right)\\
\label{recursion1b}
(m+1-s)\alpha\tilde{b}^{(m+1)}_{s}=&
  m(1+\alpha^2)(1+H^2)\tilde{b}^{(m)}_{s}-
  (m+s-1)\alpha\tilde{b}^{(m-1)}_{s},
\end{align}
\end{subequations}
can be used to simplify equations (\ref{fg1}) further. Equations
(\ref{recursion1}) are derived in \cite{BC61} for the case
where $H=0$. However the more general relations given above
are readily obtained by replacing the combination $1+\alpha^2$
appearing in the \cite{BC61} recursion relations with
$(1+\alpha^2)(1+H^2)$. So for $m=1$ and $s=3/2$ equation
(\ref{recursion1a}) is
\begin{equation}
\label{recursion2}
\tilde{b}^{(1)}_{3/2}=\frac{3}{2}\alpha
  \left(\tilde{b}^{(0)}_{5/2}-\tilde{b}^{(2)}_{5/2}\right)
\end{equation}
and
\begin{equation}
\label{recursion3}
\tilde{b}^{(2)}_{3/2}=\frac{3}{4}\alpha
  \left(\tilde{b}^{(1)}_{5/2}-\tilde{b}^{(3)}_{5/2}\right)
\end{equation}
for $m=2$ and $s=3/2$ while equation (\ref{recursion1b}) yields
\begin{equation}
\label{recursion4}
2(\alpha^2+1)(1+H^2)\tilde{b}^{(1)}_{3/2}=
  \alpha\tilde{b}^{(2)}_{3/2}+3\alpha\tilde{b}^{(0)}_{3/2}
\end{equation}
for $m=1$ and $s=3/2$. Inserting equations
(\ref{recursion2}--\ref{recursion4}) into
(\ref{fg1}) then yields
\begin{subequations}
\begin{align}
\label{fg2}
f(\alpha,\mathfrak{h},\mathfrak{h'})=&
  \alpha\tilde{b}^{(1)}_{3/2}-
  3\alpha^2H^2(2+H^2)\tilde{b}^{(0)}_{5/2}\\
g(\alpha,\mathfrak{h},\mathfrak{h'})=&
  -\alpha\tilde{b}^{(2)}_{3/2}+
  3\alpha^2H^2(2+H^2)\tilde{b}^{(1)}_{5/2}.
\end{align}
\end{subequations}

\section{Appendix \ref{appendixC}}
\label{appendixC}

The symbolic mathematics software MAPLE
has been used to write the needed softened Laplace coefficients 
$\tilde{b}^{(m)}_{s}$, equation (\ref{b_soft_int}), in term of
complete elliptic integrals $K$ and $E$. Setting
$H^2=\onehalf(\mathfrak{h}^2+\mathfrak{h'}^2)$,
$\gamma=(1+\alpha^2)(1+H^2)/2\alpha$, 
and $\chi=\sqrt{2/(\gamma+1)}$, then
\begin{subequations}
\label{b}
\begin{align}
\tilde{b}^{(0)}_{1/2}(\alpha,\mathfrak{h},\mathfrak{h'})=&
  \frac{4K(\chi)}{\pi\sqrt{2\alpha(\gamma+1)}}\\
\tilde{b}^{(1)}_{1/2}(\alpha,\mathfrak{h},\mathfrak{h'})=&
  \frac{4\left[\gamma K(\chi)-(\gamma+1)E(\chi)\right]}
  {\pi\sqrt{2\alpha(\gamma+1)}}\\
\tilde{b}^{(0)}_{3/2}(\alpha,\mathfrak{h},\mathfrak{h'})=&
  \frac{2E(\chi)}{\pi\alpha(\gamma-1)\sqrt{2\alpha(\gamma+1)}}\\
\tilde{b}^{(1)}_{3/2}(\alpha,\mathfrak{h},\mathfrak{h'})=&
  \frac{2\left[-(\gamma-1)K(\chi)+\gamma E(\chi)\right]}
  {\pi\alpha(\gamma-1)\sqrt{2\alpha(\gamma+1)}}\\
\tilde{b}^{(2)}_{3/2}(\alpha,\mathfrak{h},\mathfrak{h'})=&
  \frac{2\left[-4\gamma(\gamma-1)K(\chi)+
    (4\gamma^2-3)E(\chi)\right]}
  {\pi\alpha(\gamma-1)\sqrt{2\alpha(\gamma+1)}}\\
\tilde{b}^{(0)}_{5/2}(\alpha,\mathfrak{h},\mathfrak{h'})=&
  \frac{4\left[-(\gamma-1)K(\chi)+4\gamma E(\chi)\right]}
  {3\pi(2\alpha)^{5/2}(\gamma+1)^{3/2}(\gamma-1)^{2}}\\
\tilde{b}^{(1)}_{5/2}(\alpha,\mathfrak{h},\mathfrak{h'})=&
  \frac{4\left[-\gamma(\gamma-1)K(\chi)+
    (\gamma^2+3)E(\chi)\right]}
  {3\pi(2\alpha)^{5/2}(\gamma+1)^{3/2}(\gamma-1)^{2}}
\end{align}
\end{subequations}
where
\begin{equation}
K(\chi)=\int_0^1\frac{dt}{\sqrt{(1-t^2)(1-\chi^2t^2}}
\end{equation}
is the complete elliptic integral of the first kind and
\begin{equation}
E(\chi)=\int_0^1\sqrt{\frac{1-\chi^2t^2}{1-t^2}}dt
\end{equation}
is the complete elliptic integral of the second kind. The series
expansions for $K(\chi)$ and $E(\chi)$ given in \cite{AS70}
then permit the rapid calculation of the softened Laplace
coefficients $\tilde{b}^{(m)}_{s}$ without requiring a
numerical integration of equation (\ref{b_soft_int}). However
equations (\ref{b}) can give unreliable results for extreme
values of $\alpha$ due to numerical roundoff errors. In this
case it is preferable to factor the
$(1+\alpha^2)(1+H^2)=2\alpha\gamma$
term out of the integrand in (\ref{b_soft_int}) and expand the
denominator for the case of large $\gamma\gg1$:
\begin{subequations}
\begin{align}
\tilde{b}^{(m)}_{s}=&\frac{2}{\pi(2\alpha\gamma)^s}
  \int_0^\pi d\phi\cos(m\phi)
  \left[1+\frac{s}{\gamma}\cos\phi+
    \frac{s(s+1)}{2\gamma^2}\cos2\phi+...\right]\\
\label{b_approx}
\simeq&\frac{f_m\alpha^m}{(2\alpha\gamma)^{s+m}}
\end{align}
\end{subequations}
where $f_0=2$, $f_1=2s$, and $f_2=s(s+1)$. Equation
(\ref{b_approx}) usually gives the more reliable result for
$\alpha\ll0.01$ and for $\alpha\gg100$.

Another useful form for $\tilde{b}^{(m)}_{s}$ is obtained for
regions where $\alpha=1+x$ where $|x|\ll1$. In this `local'
approximation, the dominant contribution to the integral,
equation (\ref{b_soft_int}), occurs where $\phi\ll1$. Thus we
can set $\cos\phi\simeq1-\phi^2/2$ and $\alpha\simeq1$
except where it appears as $\alpha-1=x$, extend the upper
integration limit to infinity, and set $\cos(m\phi)\simeq1$
in the numerator [cf. \cite{GT80}]:
\begin{equation}
\label{b_local}
\tilde{b}^{(m)}_{s}(x)\simeq\frac{2}{\pi}\int_0^\infty
\frac{d\phi}{(x^2+2H^2+\phi^2)^s}=\left\{
\begin{split}
\frac{2/\pi}{x^2+2H^2}\quad\mbox{for $s=3/2$}\quad\\
\frac{4/3\pi}{(x^2+2H^2)^2}\quad\mbox{for $s=5/2$}\quad
\end{split}
\right.
\end{equation}

\section{Appendix \ref{appendixD}}
\label{appendixD}

The time derivatives of $L_e$, equation (\ref{Le}), is
\begin{subequations}
\begin{align}
\frac{dL_e}{dt}=&\sum_jm_jn_ja_j^2(h_j\frac{dh_j}{dt}+
  k_j\frac{dk_j}{dt})\\
=&\sum_j\sum_{k\ne j}m_jn_j^2a_j^2A_{jk}(h_jk_k-h_kk_j)\\
=&\sum_j\sum_{k\ne j}\frac{1}{4}
  \left(\frac{m_jm_k}{M_\odot+m_j}\right)n_j^2a_j^2g_{jk}
    (h_jk_k-h_kk_j)
\end{align}
\end{subequations}
where $g_{jk}=g(\alpha_{jk},\mathfrak{h}_j,\mathfrak{h}_k)$.
Let $dL_e/dt=S_1-S_2$ where $S_1$ is the sum over the $h_jk_k$
terms and $S_2$ the sum over the $h_kk_j$. Swap the $j$ and $k$
indices in $S_2$ so that it becomes
\begin{equation}
S_2=\sum_k\sum_{j\ne k}\frac{1}{4}
  \left(\frac{m_km_j}{M_\odot+m_k}\right)n_k^2a_k^2g_{kj}h_jk_k.
\end{equation}
Equation (\ref{g_reciprocal}) shows that
$g_{kj}=\alpha_{jk}g_{jk}$, and with
$(n_k/n_j)^2=\alpha_{jk}^{-3}(M_\odot+m_k)/(M_\odot+m_j)$,
$S_2$ becomes
\begin{equation}
S_2=\sum_k\sum_{j\ne k}\frac{1}{4}
  \left(\frac{m_jm_k}{M_\odot+m_j}\right)n_j^2a_j^2g_{jk}h_jk_k
\end{equation}
which is $S_1$ since the sums obey
$\sum_k\sum_{j\ne k}=\sum_j\sum_{k\ne j}$.
Consequently $dL_e/dt=S_1-S_2=0$, and a similar analysis will
also show that $dL_i/dt=0$.


\newpage

\begin{figure}
\epsscale{1.0}
\plotone{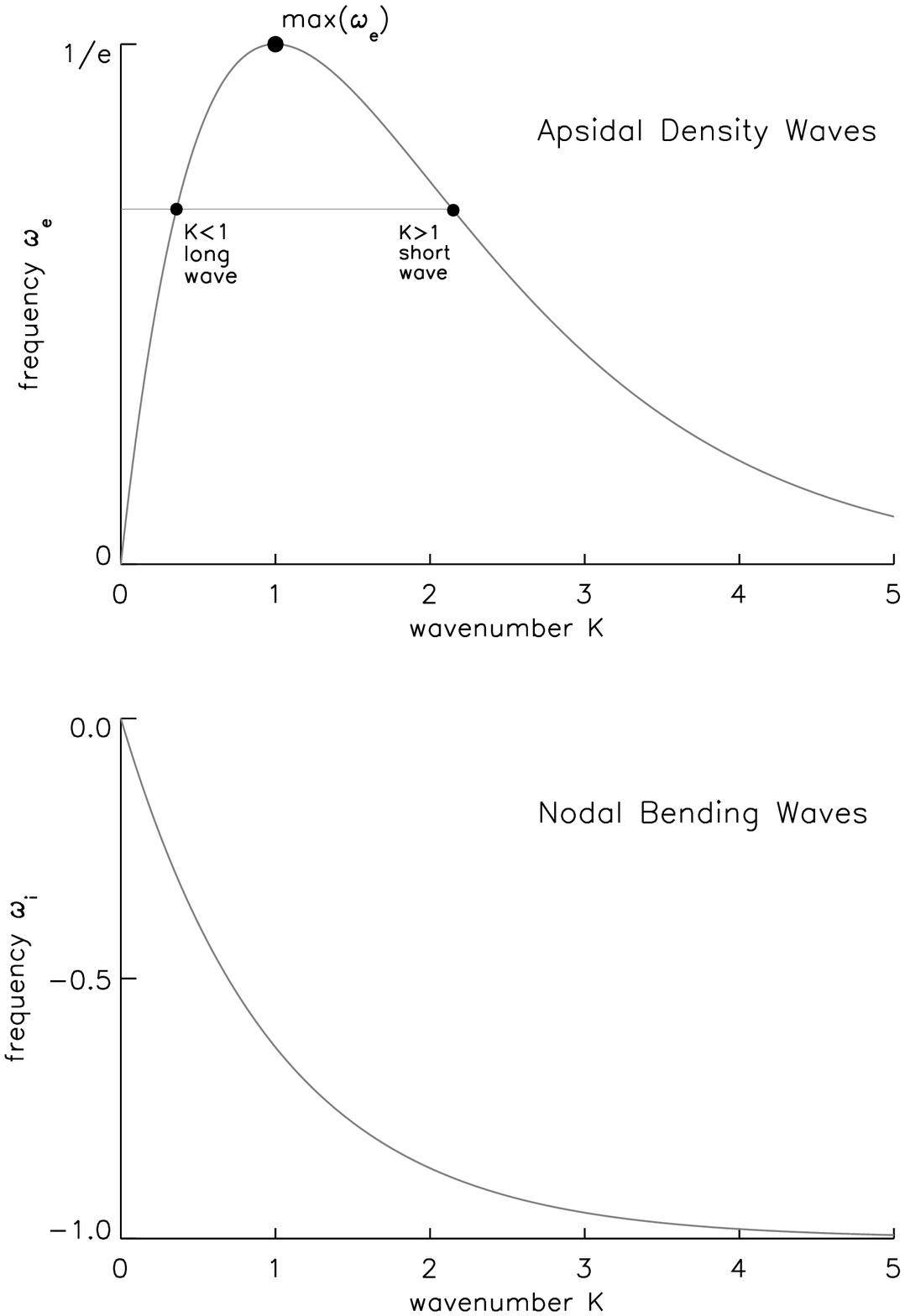}
\end{figure}

\begin{figure}
\figcaption{The upper figure is the dispersion relation
$\omega_e=Ke^{-K}$ for apsidal density waves, and the lower
figure in the dispersion relation $\omega_i=e^{-K}-1$ for
nodal bending waves.
\label{disp_rel}}
\end{figure}

\begin{figure}
\epsscale{1.0}
\vspace*{-4ex}\plotone{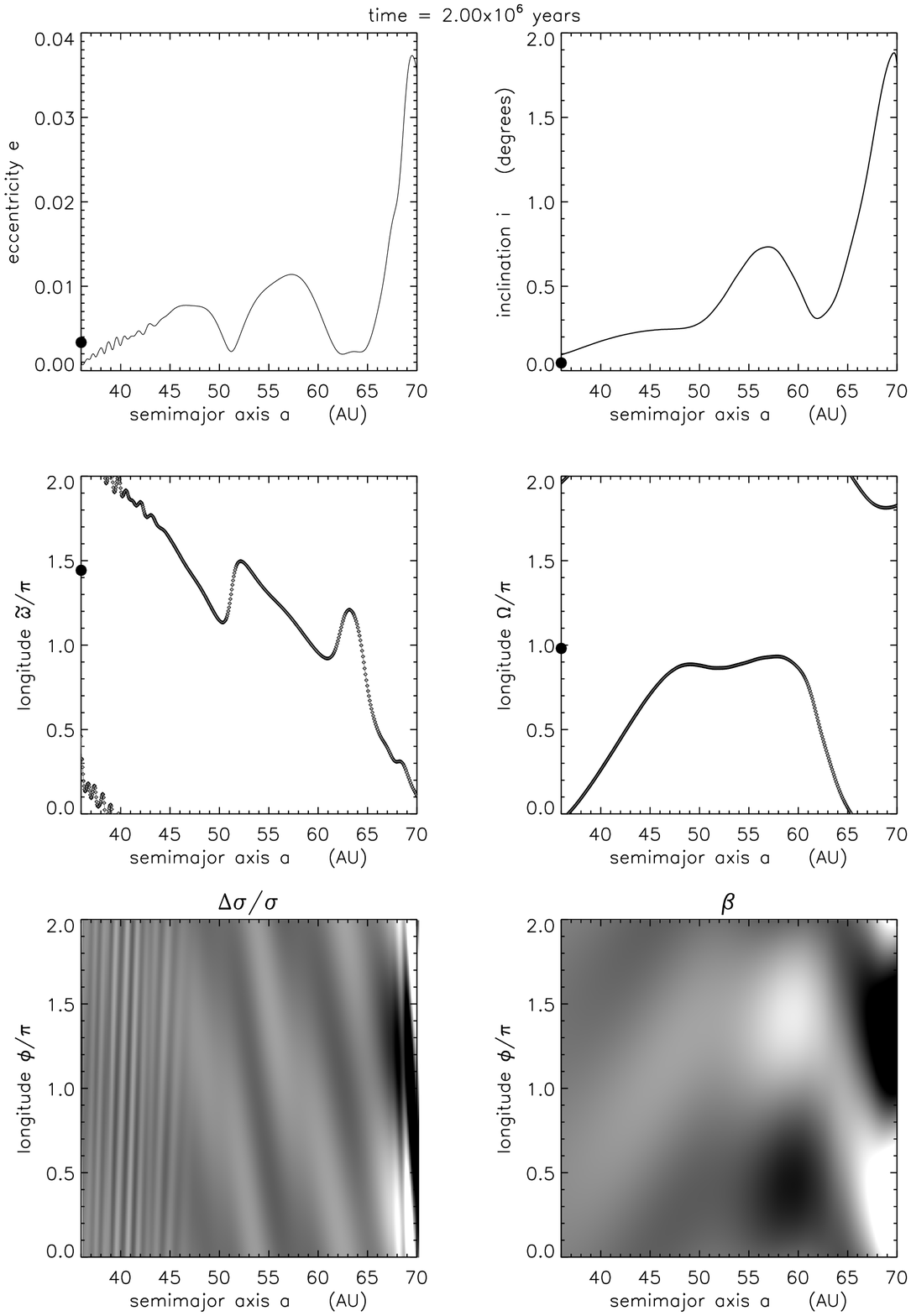}
\end{figure}

\begin{figure}
\figcaption{A snapshot of the simulated
Kuiper Belt's orbit elements
$(e,i,\tilde{\omega},\Omega)$ plotted versus semimajor axis $a$
at time $t=2\times10^6$ years in a $M_{KB}=10$ M$_\oplus$ Belt.
The Belt's fractional half--thickness is $\mathfrak{h}=0.0067$.
The dots along the left axes indicate Neptune's orbit elements.
The $\Delta\sigma/\sigma$ greyscale shows the disk's
fractional surface density variations versus
the polar coordinates $(a,\phi)$, estimated via equation
(\ref{Delta-sigma}). White/black zones indicate
over/under dense regions where $|\Delta\sigma/\sigma|$ exceeds 
0.63, while white/black zones in the other greyscale indicate the 
disk's latitude $\beta=z/a$
above/below the invariable plane, with saturation occurring
wherever $|\beta|$ exceeds $0.86^\circ$. The online edition of
this figure will be linked to an animated
sequence of these figures that show the system's time--evolution.
\label{M_KB=10}}
\end{figure}

\begin{figure}
\epsscale{1.0}
\plotone{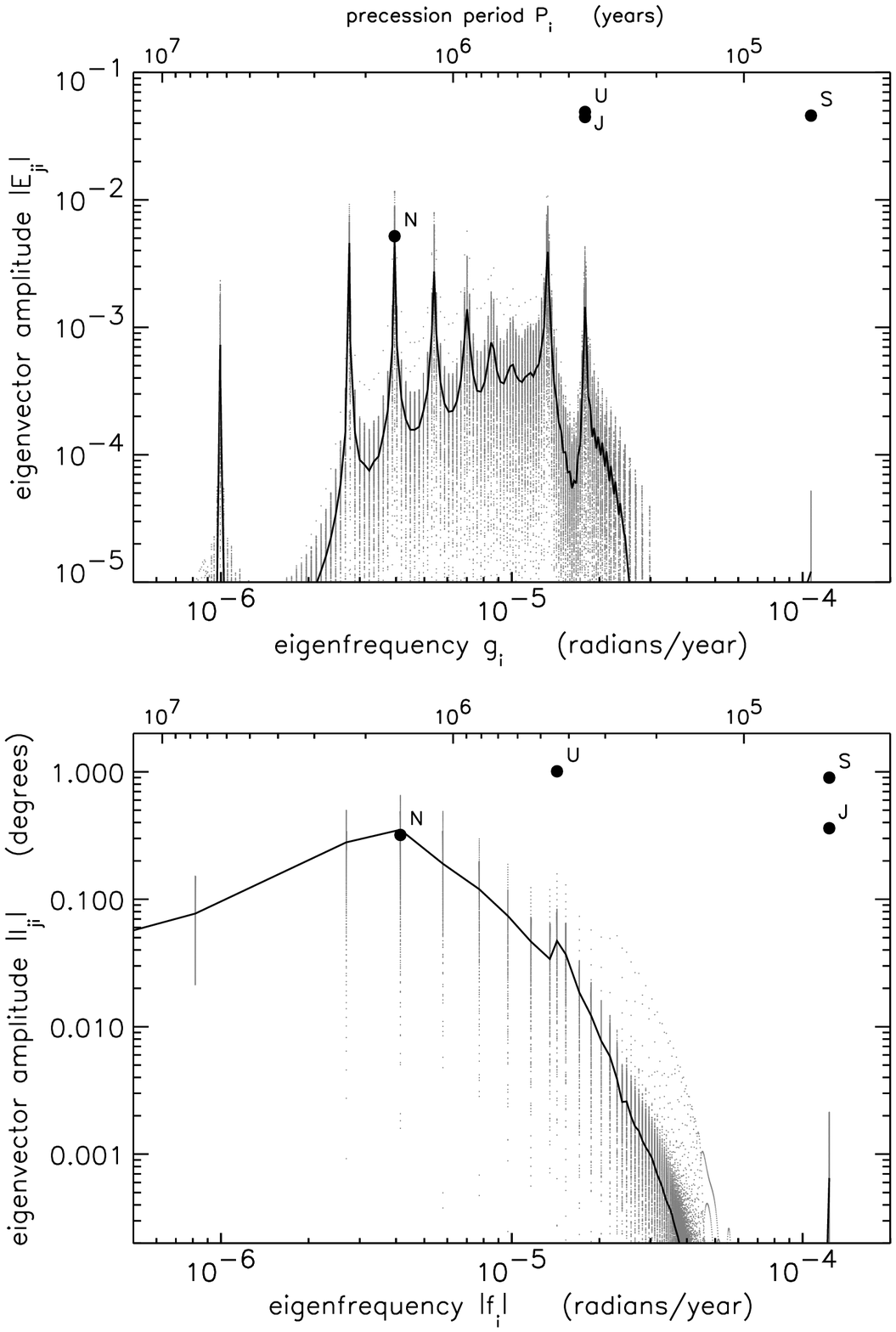}
\end{figure}

\begin{figure}
\figcaption{The numerous small dots show the individual elements
in the disk--rings' eccentricity eigenvectors $|E_{ji}|$, all
plotted versus their eigenfrequencies $g_i$, as well as the
disk--rings inclination eigenvector elements $|I_{ji}|$ versus
$|f_i|$, for the system shown in Fig.\ \ref{M_KB=10}.
The solid curves are $|E_{ji}|$ and $|I_{ji}|$ averaged
over semimajor axes $45\le a_j\le55$ AU.
The upper horizontal axes are the
corresponding precession periods $2\pi/g_i$ and $2\pi/|f_i|$.
The large dots indicate the eigenfrequency and eigenvector
element that dominates the motion of each giant planet
J=Jupiter, S=Saturn, U=Uranus, and N=Neptune.
\label{eigenvectors}}
\end{figure}

\begin{figure}
\epsscale{1.0}
\plotone{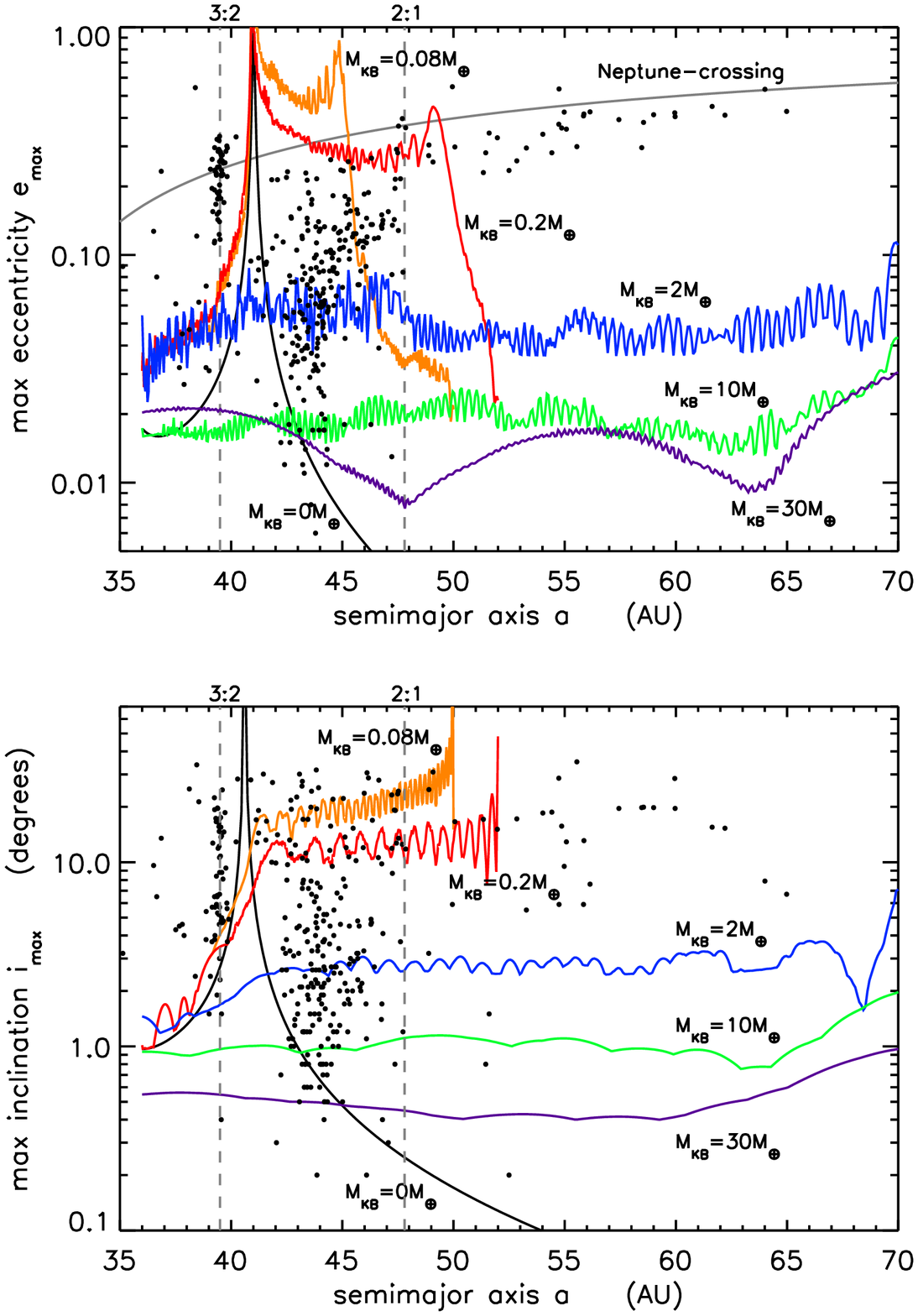}
\end{figure}

\begin{figure}
\figcaption{The maximum eccentricities
$e_{\mbox{\scriptsize max}}$ and inclinations
$i_{\mbox{\scriptsize max}}$ versus semimajor axis $a$ for
simulations having a variety of Kuiper Belt masses
M$_{KB}$; see Section \ref{mass-variations} for model details.
Dots indicate the orbits of 340 KBOs observed over multiple
oppositions, with orbits provided by the Minor Planet Center.
The locations of the $3:2$ and $2:1$ resonances are indicated,
and orbits above the grey curve are Neptune--crossing.
\label{compare_ei}}
\end{figure}

\begin{figure}
\epsscale{1.0}
\plotone{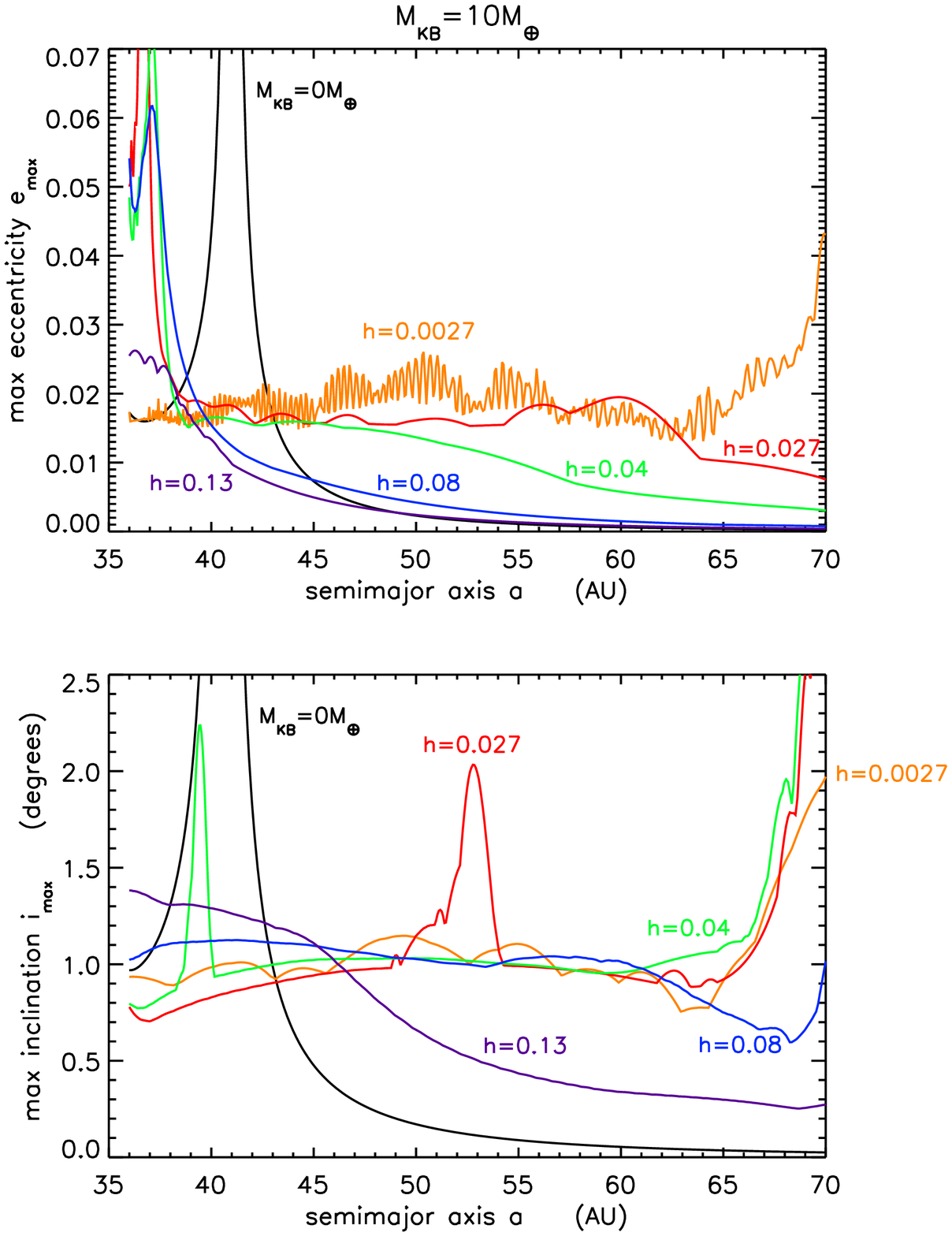}
\end{figure}

\begin{figure}
\figcaption{Maximum eccentricities and inclinations
in a $M_{KB}=10$ M$_\oplus$ disk having a fractional
thickness $\mathfrak{h}=0.0027, 0.027, 0.04, 0.08,$ and 0.13.
The black curves are the forced $e$'s and $i$'s that occur
in a massless disk. The characteristic particles size associated
corresponding to each disk thickness $\mathfrak{h}$ may be
obtained by setting particle dispersion velocities equal to
their surface escape velocity, which corresponds to particle radii
of $R\sim5000\mathfrak{h}\mbox{ km}\sim14, 140, 200, 400$, and
650 km, respectively. 
\label{M=10}}
\end{figure}

\begin{figure}
\epsscale{1.0}
\plotone{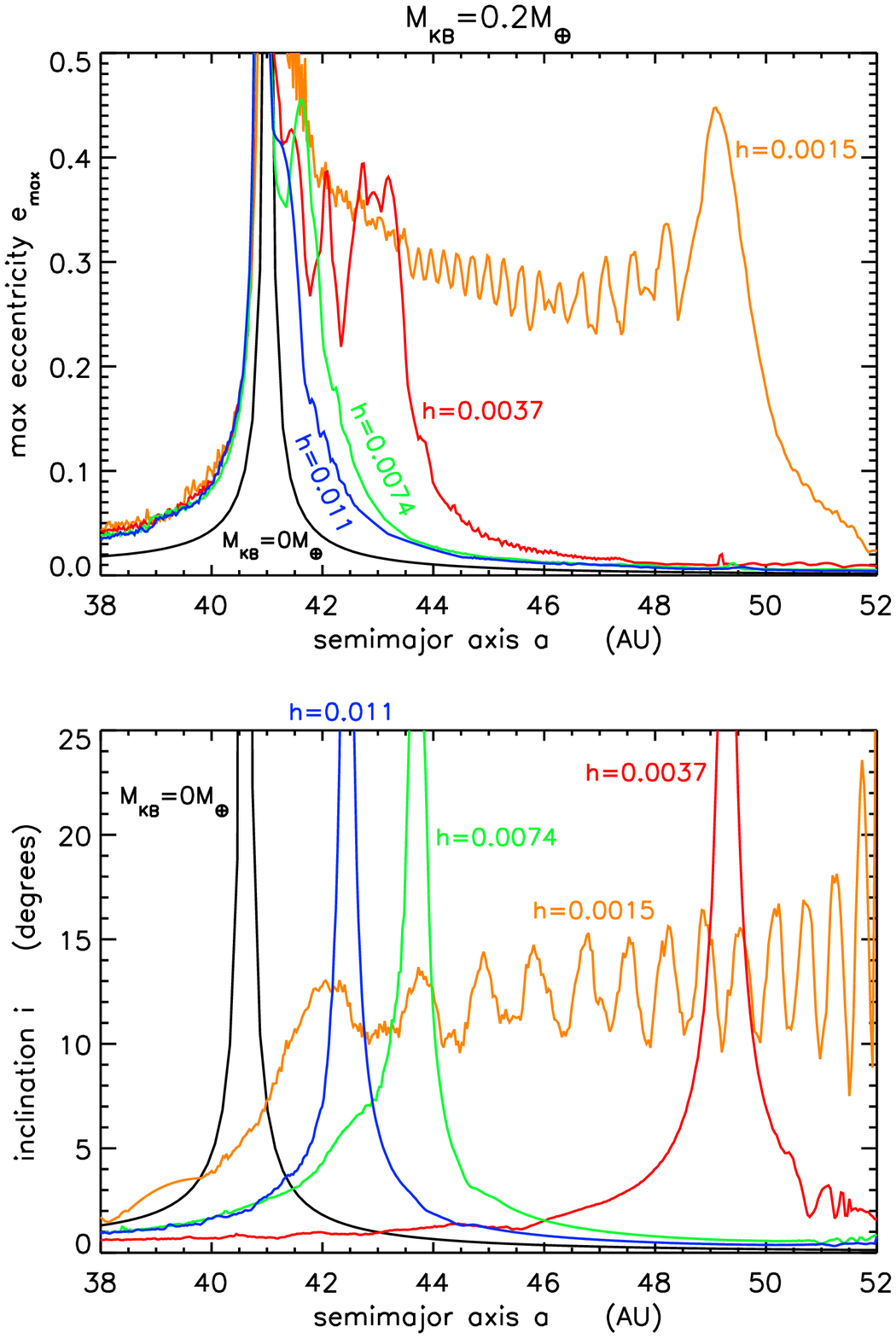}
\end{figure}

\begin{figure}
\figcaption{The upper plot gives the maximum
eccentricities $e_{\mbox{\scriptsize max}}$
in a $M_{KB}=0.2$ M$_\oplus$ disk having thicknesses
$\mathfrak{h}=0.0015, 0.0037, 0.0074,$ and 0.011. The lower plot
gives the {\sl average} inclinations $i$ that occur after the
bending wave has stalled. The exception is the
orange $\mathfrak{h}=0.0015$ curve; this wave does not stall, so
this curve shows
the maximum inclination $i_{\mbox{\scriptsize max}}$.
The black curves are the forced $e$'s and $i$'s that occur
in a massless disk. 
\label{M=0.2}}
\end{figure}

\end{document}